\documentclass[aps, prx, twocolumn, superscriptaddress, showpacs, floatfix, longbibliography, american, 10pt]{revtex4-2}

\usepackage{graphicx}
\usepackage{bm}

\usepackage{qcircuit}

\usepackage[colorlinks=true, allcolors=blue]{hyperref}

\usepackage{url}
\usepackage{amsmath}
\usepackage{amssymb}
\usepackage{braket}
\usepackage{multirow}
\usepackage[shortcuts]{extdash}

\usepackage{mathrsfs} 
\usepackage{bm} 
\usepackage{physics}
\usepackage{appendix}
\usepackage{algorithm}
\usepackage[noend]{algpseudocode}

\usepackage{enumitem}

\begin{document}

\title{Entropy-driven entanglement forging}

\author{A.~P\'{e}rez-Obiol}
\thanks{corresponding author, axel.perezobiol@uab.cat}
\affiliation{Departament de Física, Universitat Autònoma de Barcelona, 08193 Bellaterra, Spain}

\author{ S.~Masot-Llima}
\email{sergi.masot@bsc.es}
\affiliation{Barcelona Supercomputing Center, 08034 Barcelona, Spain}

\author{ A.~M.~Romero}
\email{antonio.marquezromero@fujitsu.com}
\affiliation{
Departament de Física Quàntica i Astrofísica, Universitat de Barcelona, 08028 Barcelona, Spain}
\affiliation{
Institut de Ciències del Cosmos, Universitat de Barcelona, 08028 Barcelona, Spain}

\author{ J.~Men\'{e}ndez}
\email{menendez@fqa.ub.edu}
\affiliation{
Departament de Física Quàntica i Astrofísica, Universitat de Barcelona, 08028 Barcelona, Spain}
\affiliation{
Institut de Ciències del Cosmos, Universitat de Barcelona, 08028 Barcelona, Spain}

\author{ A.~Rios}
\email{arnau.rios@fqa.ub.edu}
\affiliation{
Departament de Física Quàntica i Astrofísica, Universitat de Barcelona, 08028 Barcelona, Spain}
\affiliation{
Institut de Ciències del Cosmos, Universitat de Barcelona, 08028 Barcelona, Spain}

\author{ A.~Garc\'{i}a-S\'{a}ez }
\email{artur.garcia@bsc.es}
\affiliation{Barcelona Supercomputing Center, 08034 Barcelona, Spain}
\affiliation{Qilimanjaro Quantum Tech, 08019 Barcelona, Spain}

\author{ B.~Juli\'{a}-D\'{i}az}
\email{bruno@fqa.ub.edu}
\affiliation{
Departament de Física Quàntica i Astrofísica, Universitat de Barcelona, 08028 Barcelona, Spain}
\affiliation{
Institut de Ciències del Cosmos, Universitat de Barcelona, 08028 Barcelona, Spain}

\date{Received: \today{} / Revised version: date}

\begin{abstract}
Simulating physical systems with variational quantum algorithms is a well-studied approach, but it is challenging to implement in current devices due to demands in qubit number and circuit depth. We show how limited knowledge of the system, namely the entropy of its subsystems, its entanglement structure or certain symmetries, can be used to reduce the cost of these algorithms with entanglement forging. To do so, we simulate a Fermi-Hubbard one-dimensional chain with a parametrized hopping term, as well as atomic nuclei ${}^{28}$Ne and ${}^{60}$Ti with the nuclear shell model. Using an adaptive variational quantum eigensolver we find significant reductions in both the maximum number of qubits (up to one fourth) and the amount of two-qubit gates (over an order of magnitude) required in the quantum circuits. Our findings indicate that our method, entropy-driven entanglement forging, can be used to adjust quantum simulations to the limitations of noisy intermediate-scale quantum devices.

\end{abstract}

\maketitle

\section{Introduction}

In modern quantum devices, the number of available qubits and low-error quantum gates imposes a strong limitation in the accuracy of the final results not only in fault-tolerant schemes but also in current noisy simulations. 
Since quantum algorithms are applied to complex quantum many-body problems~\cite{Cerezo2021,Tilly2022}, including quantum chemistry~\cite{uccrev,McArdle2020,haidar2022open}, condensed matter ~\cite{cade2020strategies,perezobiol_2022,tang2024} and nuclear physics~\cite{Dumitrescu2018,Ayral:2023ron, Watson:2023oov},
this constraint has been the source of many techniques that aim to simulate large systems with smaller but equivalent ones needing fewer quantum resources. Some encoding methods aim to reduce the dimension of the input data, like quantum autoencoding~\cite{Romero_2017} or other physically-inspired frameworks~\cite{Parella-Dilme:2023qps}. On the other hand, other approaches focus directly on circuits to run equivalent simulations using fewer qubits. The best-known of them, circuit knitting~\cite{PhysRevX.6.021043,10236453}, takes advantage of circuits with sparsely-connected subsections in order to break them apart into smaller ones. More recently, an alternative approach has also been explored: one can train small circuits to prepare local states, and then recover the global solution with post-processing. Both types of techniques are commonly used on variational quantum algorithms (VQA) \cite{Cerezo2021,Tilly2022}. 

Entanglement forging~\cite{PRXQuantum.3.010309} is an example of the latter approach. However, a challenge faced by this framework is to find which terms and corresponding weights are more relevant in the entanglement-forging decomposition. Recently, a solution was proposed using a generative neural network~\cite{PhysRevResearch.6.023021}. Instead, in this work we argue in favour of a physically-motivated approach. In particular, we aim to exploit information about the entanglement structure of the target system~\cite{Klco:2021lap} or, more generally, the entropy of its subsystems and certain symmetries of the Hamiltonian. These properties can indicate how to apply the decomposition and guide the weight distribution of the product states. 

\begin{figure*}[t!]
     \centering
     \includegraphics[width=\linewidth]{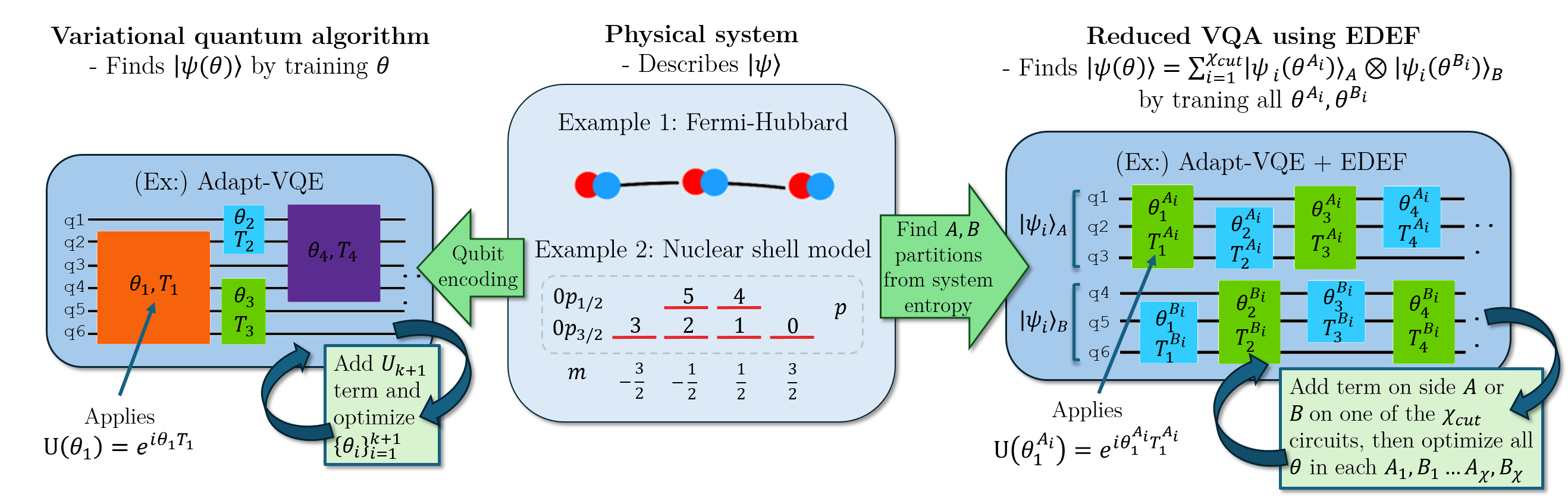}
     \caption{Schematic application of the Entropy-Driven Entanglement Forging (EDEF) method to two physical systems: the Fermi-Hubbard and the nuclear shell model. In the middle, the initial configuration is encoded into a qubit quantum state. Then --left panel-- a variational quantum algorithm (VQA) can be used to obtain the system's ground-state by optimizing the ansatz parameters. Alternatively (right panel) the VQA is used with EDEF: knowledge of the entropy of the system is used to find a suitable partition, and thus the ground state is obtained with a few smaller circuits instead of a single large circuit.
     } 
     \label{fig:summary}     
\end{figure*}

We call this novel approach Entropy-Driven Entanglement Forging (EDEF). Figure \ref{fig:summary} illustrates the EDEF algorithm, which can be applied to a physical system (middle panel) that, once encoded into qubits, can be solved with a VQA (left panel). When two low-entanglement subsystems $A$ and $B$ can be identified, they define a bipartition of the system where entanglement forging can be applied efficiently (right panel). In this alternative approach, the VQA is simplified into smaller circuits with fewer qubits that output local states for the $A$ and $B$ subsystems after optimization. Finally, we recover the ground state of the entire system with a linear combination of these local states. 

We showcase our proposal studying two many-body systems. First, we use the one-dimensional (1D) Fermi-Hubbard (FH) model. Variational \cite{cade2020strategies,Consiglio_2022,Stanisic2022,Gyawali22,tang2024} and adiabatic \cite{perezobiol_2022} quantum algorithms have been devised and tested to find ground states of the FH Hamiltonian. These algorithms have focused mainly on optimizing circuit depths and on simulations of small FH lattices given the current limitations on the number of qubits per chip. 
Here we explore the performance of the EDEF algorithm across a range of values for the different parameters of the Hamiltonian, which lets us identify partitions with varying levels of entropy. Second, we simulate the ground states of atomic nuclei using the nuclear shell model (NSM). This approach has already been the subject of several works using quantum computing tools \cite{romeroquantum,papenbrock,stetcu,Sarma:2023aim,Bhoy:2024uuh,Yoshida:2024ubi,Li:2023eyg,PhysRevResearch.6.023021}, which point out the need for a significant amount of quantum-computing resources to simulate medium-mass nuclei~\cite{PerezMarquez2023,Li:2023eyg}. 
For both systems, we implement EDEF onto an Adaptive Derivative-Assembled Pseudo-Trotter ansatz-Variational Quantum Eigensolver (ADAPT-VQE), which has previously shown to be effective both for the FH model \cite{Gyawali22} and the NSM~\cite{romeroquantum,PerezMarquez2023}. Nonetheless, our approach can be integrated with other VQAs in a straightforward way. 
As demonstrated in the following sections, the use of EDEF effectively reduces both the number of qubits and the circuit depth in our FH and NSM many-body simulations. This is particularly beneficial for noisy intermediate-scale quantum devices and holds promise for advancing the broader challenge of simulating quantum many-body systems.

\section{Entanglement patterns and physical models}\label{physical_models}

\subsection{Entropy-driven entanglement forging}

For any possible bipartition of a quantum system, bipartite entanglement \cite{horodecki} quantifies how correlated the two parts are -- albeit with some subtleties that distinguish quantum effects from classical correlations \cite{doi:10.1142/9789813237230_0007}. In a pure quantum state, $\ket{\psi}$, with density matrix, $\rho = \ket{\psi}\bra{\psi}$, subsystems $A$ and $B$ of $\rho$ are not entangled when they can be written down as a tensor product, $\rho = \rho_A \otimes \rho_B$. Otherwise, one can use the von Neumann entropy, $S$, defined as
\begin{equation}\label{eq:vnentropy}
    S(\rho) = - \text{Tr}(\rho \log_2 \rho) = - \sum_i \rho_i \log_2 \rho_i,
\end{equation}
where $\rho_i$ are the eigenvalues of $\rho$. Specifically, for a pure state, $S(\ket{\psi}\bra{\psi})=0$, while for a reduced matrix $\rho_A=tr_B(\rho)$, $S(\rho_A)=S(\rho_B)=S(tr_A(\rho))$ quantifies the entanglement between $A$ and $B$. 

The Schmidt decomposition of a quantum state \cite{Nielsen_Chuang_2010} for the same bipartition $A$,$B$,
\begin{equation}
\label{eq:schmidt}
    \ket{\psi} = \sum_i^\chi \lambda_i \ket{\psi_i}_A\otimes\ket{\psi_i}_B,
\end{equation}
is useful to calculate the entropy because $\rho_i = \lambda_i^2$. More importantly, it also describes exactly how to assemble in entanglement forging the subcircuits that describe the two parts, $\ket{\psi_i}_A$ and $\ket{\psi_i}_B$, into the full state, $\ket{\psi}$. For an exact simulation, the amount of subcircuits needed corresponds to the Schmidt number $\chi$, which is bound by the amount of basis elements in the smallest partition. On the other hand, $\chi$ is the number of terms in the sum of Eq.~(\ref{eq:vnentropy}), which maximizes the entropy when all $\rho_i$ are equal and $\Tr(\rho)=1$, with $S=\log_2(\chi)$. Therefore, 
for an equipartite $N_q$-qubit system such as the ones we consider in this work, we have
\begin{equation}\label{eq:ent_bound}
    2^S \leq \chi \leq 2^{N_q/2}.
\end{equation}

Quantum systems that are fully separable on a bipartition only need one state for each part, therefore, entanglement forging is simplest for these systems. For low entanglement between the two parts, only a few instances $\chi_{\rm cut}$ of each subsystem are necessary to simulate it accurately, since there is a tail of terms with very small coefficients $\lambda_i$ that contribute negligibly. In contrast, for strongly-entangled subsystems one needs exponentially many states with the smallest number of qubits in one of the subsystems. Thus, in a general setting of entanglement forging, one must find a favourable bipartition, decide how many subcircuits to run, and optimize the coefficients for each of them to recover the full quantum state.
Since the number of possible bipartitions in a system scales exponentially with the number of qubits $N_q$, entanglement forging is best suited when it is physically driven, for instance when knowledge of the system indicates low entanglement across specific sectors.

In this work, we illustrate EDEF by studying the 1D FH model with a tunable hopping term between the two central sites. This model allows us to test different levels of entropy and their impact on the quality of the results after applying one layer of EDEF.
In addition, the low entanglement between protons and neutrons in atomic nuclei, as demonstrated with the NSM in Refs.~\cite{johnson2023proton,perezobiol2023}, provides an ideal practical testbed for EDEF. While we focus on the ADAPT-VQE algorithm, other VQAs can also be applied to the local circuits. We therefore expect EDEF to be algorithm-agnostic.

For a general physical system, we define Entropy-Driven Entanglement Forging as the following algorithm: 
\renewcommand{\thealgorithm}{}
\begin{algorithm}[H]
\caption{Entropy-Driven Entanglement Forging}\label{EDEF}
\begin{algorithmic}[1]
    \Procedure{Preparation}{}
        \For {each of \textit{l} layers of EDEF}
            \State Identify low entropy bipartition \textit{A,B}
            \State Define initial basis of states $\ket{\psi_i^0}_A,\ket{\psi_i^0}_B$
            \State $\text{Tie degenerate coefficients } \lambda_i \text{\,using symmetry}$
            \State Set cutoff number of product states $\chi_{\rm cut}$
        \EndFor
    \EndProcedure
    \Procedure{Training }{}
        \State Define state $\ket{\psi} = \sum_i^{\chi_{\rm cut}} \lambda_i \ket{\psi_i^0}_A \otimes \ket{\psi_i^0}_B$
        \While {$\varepsilon' > $ \textit{threshold}}
            \State $\text{Optimize }\mathcal{U}^i_A(\theta), \mathcal{U}^i_B(\theta) \text{ independently on circ. } i$
            \State Compute $\mathcal{U}_A \bigotimes \mathcal{U}_B\to\mathcal{U}_{VQA}$ 
            \State Compute $\varepsilon'(\mathcal{U}_{VQA},\ket{\psi})$
        \EndWhile
    \EndProcedure
\end{algorithmic}
\end{algorithm}
\noindent
In this pseudocode, $A$, $B$ denote the two partitions of the system, the unitaries $\mathcal{U}^i_A$, $\mathcal{U}^i_B$ are applied individually to circuit $i$ of the decomposition on the corresponding partition, and $\varepsilon'$ is the ground-state energy found with EDEF. The \textit{threshold} to determine the stopping point depends on the context of application, and it can be motivated by a physical purpose or by the limitations of the algorithm. The bases in the fourth step must have well-defined quantum numbers according to the Hamiltonian and the partition. Throughout this work, we refer to the application of $l$ layers of EDEF as $l$-cut EDEF.

\subsection{Fermi-Hubbard model}
\label{sec:fh}

The FH model describes fermions on a lattice and serves as a simplified model to simulate valence electrons on a crystal \cite{fermi-hubbard-model} and fermionic ultracold gases in optical lattices \cite{tarruell_fh_2018}. In its simplest 1D form, the FH Hamiltonian includes a hopping term, $t$, accounting for the tunneling of fermions between adjacent sites, and an interaction term, $U$, which adds energy whenever a spin-up and a spin-down fermion occupy the same site.
Here we consider a 1D lattice with an even number of sites, labeled $i=1,\cdots,N_s$, and a tunable hopping, $t_m$, between the two central sites, $i_m=N_s/2$ and $j_m=i_m+1$. 
The Hamiltonian reads,
\begin{equation}
\label{eq:hfh}
\begin{aligned}
    H_0 =& -t \sum_{\langle i,j \rangle,\sigma} \left( a^\dagger_{i\sigma} a_{j\sigma} + a^\dagger_{j\sigma} a_{i\sigma} \right) + U\sum_i n_{i\uparrow} n_{i\downarrow}
\\ &-(t_m-t)\sum_\sigma\left( a^\dagger_{i_m\sigma} a_{j_m\sigma} + a^\dagger_{j_m\sigma} a_{i_m\sigma} \right),
\end{aligned}
\end{equation}
where $a_{i \sigma}$ ($a^\dagger_{i \sigma}$) are the annihilation (creation) operators for a fermion at site $i$ and spin $\sigma$, with $i=1,\cdots,N_s$ and $\sigma=\uparrow,\downarrow$; $n_{i\sigma}=a^\dagger_{i \sigma} a_{i \sigma}$ is the number operator, while $\langle i,j \rangle$ indicates pairs of first-neighbor sites.
We consider repulsive interactions, $U>0$, with $t>0$, $t_m>0$.
Here, we set $t$ as our energy unit and vary $t_m$ and $U$, while fixing the number of particles for each spin, $N_\sigma$.
\begin{figure}[t]
     \centering
     \includegraphics[width=0.96\linewidth]{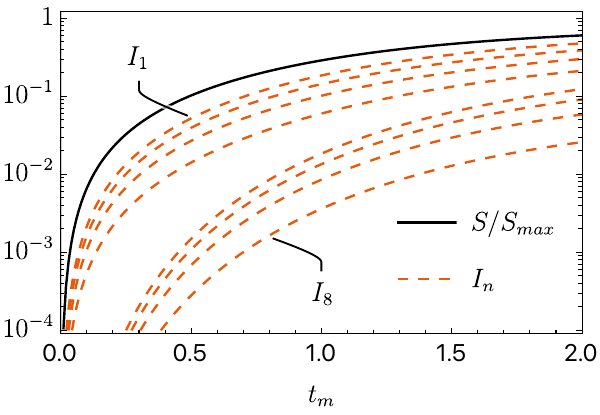}
     \caption{Normalized entropy $S/S_{max}$ (black solid line) and infidelities $I_n$ corresponding to normalized Schmidt decompositions with $n=1,\cdots,8$ product states (dashed lines, from top to bottom) as a function of the hopping $t_m$ between the two middle sites for a FH model with $N_s=4$ sites and $U=t$.
     } 
     \label{fig:fh_infs}
\end{figure}
The regular FH Hamiltonian corresponds to $t_m=t$, while smaller (larger) values of $t_m$ couple more weakly (strongly) the left and right partitions of the lattice, thus reducing (enhancing) their entanglement. In a standard Jordan-Wigner mapping~\cite{jw} with $N_q=2N_s$ qubits, odd qubits ($i=1,3,\cdots,2N_s-1$) correspond to spin-up sites and even qubits ($i=2,4,\cdots,2N_s$) to spin-down sites. The entanglement is then given by the von Neumann entropy in Eq.~(\ref{eq:vnentropy}), $S$, between the first and second halves of qubits, and following Eq.~(\ref{eq:ent_bound}) it is upper bounded by $S_{max}=N_s$.

Figure~\ref{fig:fh_infs} shows $S/S_{max}$ as a function of $t_m$ for a lattice with $N_s=4$ sites, $U=t$, and half-filling, $N_\uparrow=N_\downarrow=2$. The entropy at $t_m=0$ vanishes as the Hamiltonian can be written as a tensor product of two parts involving only the left or right partitions with $N_s/2$ sites each. As $t_m$ increases, the entropy grows up to $S/S_{max}=0.60$ at $t_m=2t$. For larger $t_m$ (not shown in Fig.~\ref{fig:fh_infs}), the entropy keeps growing and then slightly decreases, converging to $S/S_{max}=0.75$ as $t_m\to\infty$.

The entropy can also be written in terms of Schmidt coefficients $\lambda_i$ as in Eq.~(\ref{eq:schmidt}),
\begin{equation}
S=-\sum_i \lambda_i^2 \log_2(\lambda_i^2) \, .
\end{equation}
Related to this expression, we can compute the infidelity, $I_n$, of a Schmidt decomposition cut off at the $n$-th singular value and normalized to one, $|\psi_n\rangle$, when compared to the exact ground state, $|\psi_\chi\rangle$, 
\begin{align}
    I_n =& 1-|\langle \psi_\chi |\psi_n\rangle|^2
     = 
    1-\sum_i^n \lambda_i^2.
\end{align}
This infidelity allows us to quantify the similarity between a state that has been cut at the $n$-th singular value and the complete, uncut state. It thus provides a useful proxy for the quality of the Schmidt decomposition of a given state into partitions. If only a few states from each partition are relevant, the infidelity should tend to zero rapidly as a function of $n$. In fact, $I_n$ always decreases as $n$ increases, since the singular values $\lambda_i^2$ are non-negative real numbers sorted in decreasing order. The rate of decrease as a function of $n$ is expected to depend on the entanglement structure of the system. 

Figure~\ref{fig:fh_infs} shows the infidelities for $n=1,\cdots,8$ (dashed lines) as a function of $t_m$, for the FH model described above. For $t_m=0$, there is only one singular value, so that $\lambda_1=1$ and $I_n=0$ for all values of $n$. As $t_m$ increases, the different infidelities $I_n$ grow at different rates. This indicates that, depending on the target infidelity and $t_m$ value, a different cutoff $\chi_{\rm cut}=n$ is needed to achieve a good quality description of the complete state $|\psi_\chi\rangle$. If a given value of $I_n$ is good enough for our purpose, the figure indicates which EDEF with $n=\chi_{\rm cut}$ provides a suitable approximation. For instance, in the physical case where $t_m=1$, a $1\%$ infidelity is reached with $\chi_{\rm cut}=5$. 
We note, in fact, that there is a large gap between $I_4$ and $I_5$ for values of the central hopping $t_m\lesssim t$. In fact, 
this gap guarantees that for $\chi_{\rm cut}=5$ the infidelity is lower than $10 \%$ across all values of $t_m$.

\begin{figure}[t]
     \centering
     \includegraphics[width=1\linewidth]{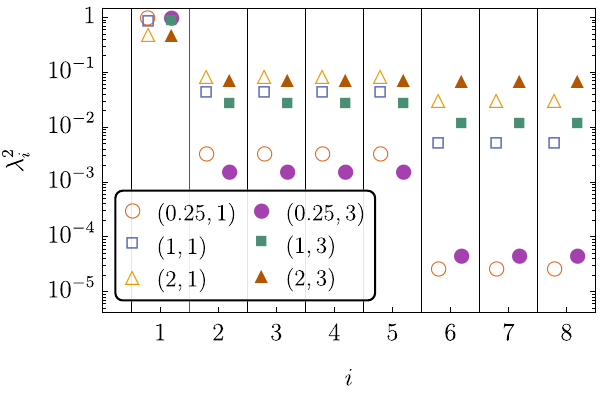}
     \caption{First eight singular values, $\lambda_i$, of the Schmidt decomposition into left and right partitions of the ground state of the FH model with 4 particles, for different interaction strengths, $U$, and central hopping terms, $t_m$, labeled ($t_m$, $U$). Empty (solid) symbols indicate singular values for an interaction strength $U=t$ ($U=3t$). As the central hopping is reduced, the values $i\geq 2$ get smaller.
     } 
     \label{fig:svs_FH}     
\end{figure}

The presence of a gap in the infidelities reflects a significant level of structure in the entanglement properties the system. 
For the FH model, the appearance of the gap can be understood in terms of the spin and parity symmetries of the FH Hamiltonian. Equation~(\ref{eq:hfh}) is invariant under the exchange of spin-up and spin-down operators, $\sigma\leftrightarrow\bar{\sigma}$, and under the exchange of each operator acting on site $i$ by the mirror operator with $i\leftrightarrow N_s+1-i$. Therefore, product states in the Schmidt decomposition related by these transformations have degenerate singular values. Figure~\ref{fig:svs_FH} illustrates this degeneracy by showing the first eight singular values for three different central hopping values, $t_m=t/2,t,2t$, and two different interactions, $U=t,3t$.
In all cases, a first large, non-degenerate, singular value is followed by four small degenerate ones, while the sixth singular value is very suppressed.
The first singular value corresponds to an even distribution of particles and spins between the left and right partitions, which we label $(\uparrow\downarrow,\uparrow\downarrow)$. For this symmetric distribution, interactions minimize the double occupation of the same site, and the hopping term delocalizes fermions. The next four degenerate singular values correspond to the partitions related by spin and parity transformations: $(\uparrow\uparrow\downarrow,\downarrow)$,
$(\downarrow,\uparrow\uparrow\downarrow)$,
$(\downarrow\downarrow\uparrow,\uparrow)$,
and $(\uparrow,\downarrow\downarrow\uparrow)$. Degenerate Schmidt coefficients decrease the infidelity by a more or less constant amount 
associated to the degenerate eigenvalues $\lambda_{2-5}$. 
The gap between $I_4$ and $I_5$ in Fig.~\ref{fig:fh_infs} appears because once the fifth product state is added, the sum of singular values suddenly approaches unity. This indicates that $n=5$ Schmidt vectors describe the vast majority of the target state. 
The following singular values are much smaller, making the corresponding infidelities relatively close to $I_5$. 

Figure~\ref{fig:fh_infs} indicates that the appearance of a gap in the infidelity is more prominent as $t_m$ decreases. 
This is somewhat natural, in that a weaker central hopping leads to a lower entanglement between left and right partitions and  
makes the $n=5$ cutoff more efficient. This is also clearly illustrated in the Schmidt coefficients shown in Fig.~\ref{fig:svs_FH}: the
lower the value of $t_m$, the smaller the Schmidt coefficients are with $i\geq 2$. 
While the infidelity results in Fig.~\ref{fig:fh_infs} are shown for $U=t$, 
different values of $U$ do not change notably the picture, as can be already anticipated from the structure of the singular values shown 
in Fig.~\ref{fig:svs_FH}. We find indeed that there is only a small difference in the Schmidt coefficients of systems with different values
of $U$, corresponding to either empty ($U=t$) or solid ($U=3t$) symbols. 

\begin{figure*}[t]
     \centering
     \includegraphics[width=0.86\linewidth]{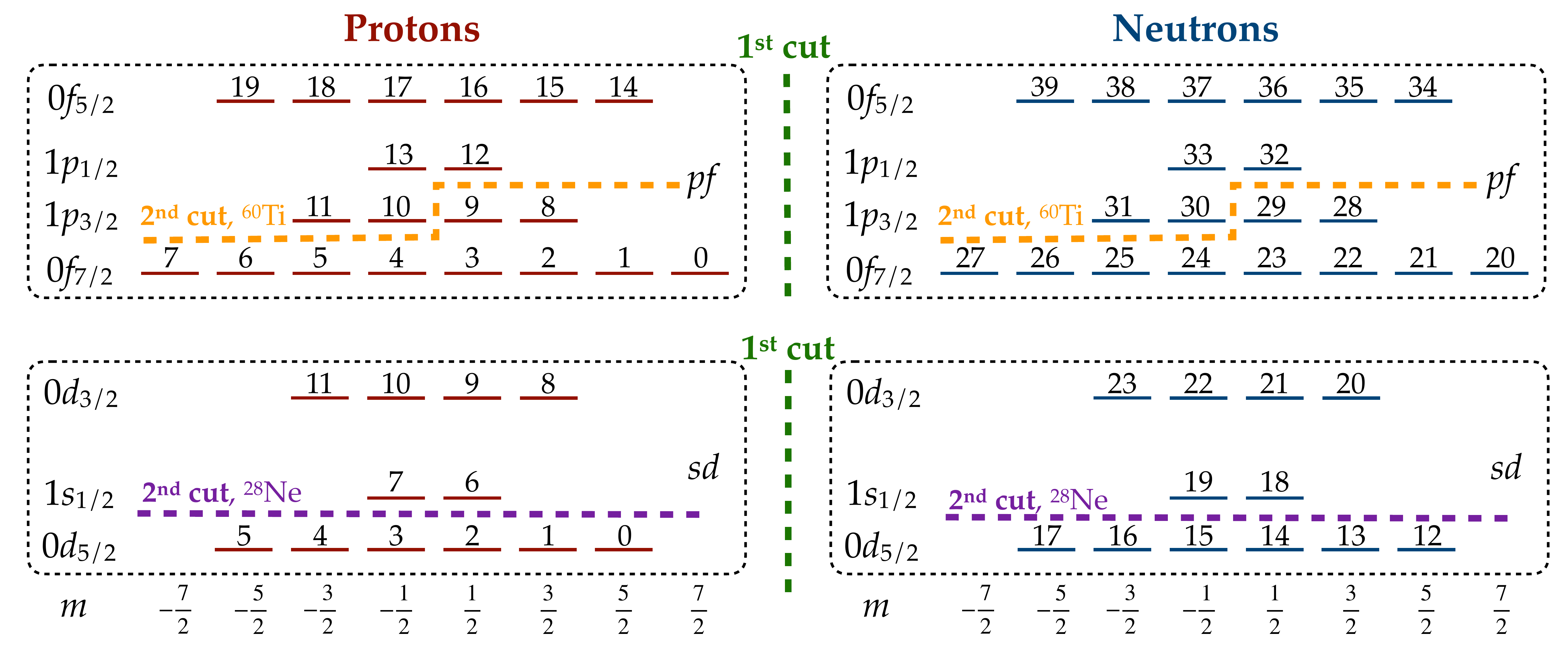}
     \caption{Partitions used in the first and second layers of EDEF (labelled as 1$^{\rm{st}}$ and 2$^{\rm{nd}}$ cut) for the nuclear shell model in the $sd$ (bottom panels) and $pf$ (top panels) valence spaces. Single-particle states are ordered according to their single-particle energy, with more deeply bound states at the bottom. The number on top of each state labels the qubit in our implementation of ADAPT-VQE under a Jordan–Wigner mapping. The first EDEF cut always separates the proton and neutron sectors, while the second cut aims to separate orbitals according to their single-particle energies. 
          } 
     \label{fig:shells}     
\end{figure*}

In the following, we fix the number of sites to $N_s=4$ and use the same set of central hopping terms $t_m$ and interactions $U$ of
Fig.~\ref{fig:svs_FH}
to illustrate the performance of EDEF in different settings.  Nonetheless, the distribution of spin-up and spin-down fermions discussed in this section, $N_\uparrow$ and $N_\downarrow$, can be generalized to lattices with more sites ({\it eg} $N_s=6$, $N_s=8$, etc) and particles, as long as interactions $U$ are not large ({\it eg} $U=t$) and $N_\uparrow=N_\downarrow$ is even. In these cases, the Schmidt decomposition has the same structure as the one presented here. A first product state consists of $N_\uparrow/2$ and $N_\downarrow/2$ spins in each side.  The following four product states are four-fold degenerate, corresponding to $N_\uparrow/2-1$ and $N_\downarrow/2+1$ on the left, and the corresponding degenerate product states obtained by left/right and spin-up/spin-down exchanges. In this sense, we expect the conclusions
that we draw with our setup to be relatively general. 

\subsection{Nuclear shell model}
\label{sec:nsm}

The nuclear shell model (NSM), or configuration-interaction method, is one of the most successful frameworks to study nuclear structure~\cite{de2013nuclear,heyde1994nuclear,brown1988status,caurier2005shell,otsuka2020evolution,Stroberg:2019mxo}. Much alike its atomic counterpart, 
the NSM characterizes nuclear dynamics in a restricted configuration space, also called valence space, where nucleons effectively interact. The valence space is bounded by single-particle states which, if completely filled with nucleons, lead to \emph{magic numbers} that characterize especially stable configurations associated with large single-particle energy gaps. As the nuclear interaction is rotationally invariant and nucleons are fermions, the single-particle basis states are labelled by the quantum numbers $pl_j$ and $m$, where $p$ is the principal quantum number; $l$, the orbital angular momentum -- usually given in spectroscopic notation --; and $j$, the total angular momentum with third-component projection $m$. As illustrated in Fig.~\ref{fig:shells}, single-particle states with the same $pl_j$ and different $m$ are degenerate in energy.

The effective Hamiltonian in the valence space is
\begin{align}\label{eq:smham}
    H_{\rm{eff}} = \sum_i \varepsilon_i a_i^{\dag} a_i + \frac{1}{4} \sum_{ijkl}\bar{v}_{ijkl}
    a_i^{\dag} a_j^{\dag} a_l a_k,
\end{align}
where the operators $a_i$ ($a_i^{\dag}$) annihilate (create) a nucleon in the single-particle state $i$ with energy $\varepsilon_i$. The antisymmetrized two-body matrix elements $\bar{v}_{ijkl} = v_{ijkl} - v_{ijlk}$ can be obtained from the full-space nucleon-nucleon interaction, but are customarily fit to specific nuclear shells. 
In this work, we use the standard USDB interaction~\cite{Brown2006} in the \emph{sd} shell for neon, and KB3G~\cite{Poves2001} in the \emph{pf} shell for titanium. Figure~\ref{fig:shells} shows the orbitals comprising these valence spaces. The bottom panels show the $0d_{5/2}$, $1s_{1/2}$ and $0d_{3/2}$ states for the $sd$ shell for protons (left) and neutrons (right). The top panels display the $0f_{7/2}$, $1p_{3/2}$, $1p_{1/2}$ and $0f_{5/2}$ states for the $pf$ shell. The vertical spacing provides an indication of the different $\varepsilon_i$ values. 
The numbers correspond to the Jordan-Wigner mapping of the different states~\cite{PerezMarquez2023,perezobiol2023}. 

For a nucleus with $Z$ protons and $N$ neutrons, one can expand the nuclear states in the many-body basis of the $M$-scheme, or Slater determinants,
\begin{equation}\label{eq:Mscheme}
|J J_z T T_z\rangle = \sum_\alpha c_\alpha |\alpha; J_z T_z\rangle,
\end{equation}
where $J_z$ is the projection of the total angular momentum of the nucleus, $J$, and $T_z = (N-Z)/2$ is the projection of the total nuclear isospin, $T$. The coefficients $c_\alpha$ are obtained by diagonalizing the Hamiltonian in the many-body basis and guarantee that nuclear states have good $J$ and $T$ quantum numbers. State-of-the-art nuclear shell-model codes ~\cite{Shimizu:2019xcd,caurier1999antoine,brown2014shell,johnson2018bigstick} face a significant challenge to build and diagonalize the Hamiltonian matrix for heavy nuclei. This is because the many-body basis, formed by the exponentially-growing set of all possible configurations for protons and neutrons in the valence space, becomes untractable for classical computers when nucleons fill about half the valence-space single-particle states. While this is not an issue in small configurations spaces like the $p$ or the $sd$ shells, current classical simulations are significantly limited in regions of the nuclide chart with higher mass numbers. 

With the advent of quantum computers, alternative quantum algorithms have been proposed to find ground states within a NSM framework~\cite{stetcu,papenbrock,Yoshida:2024ubi,PerezMarquez2023,Bhoy:2024uuh,Sarma:2023aim,PhysRevResearch.6.023021,romeroquantum, Li:2023eyg}. In parallel, recent works have provided relevant nuclear structure insights by analyzing nuclei in terms of quantum information measures such as the von Neumann entropy~\cite{Robin2021,hengstenberg2023multi,Robin:2023pgi,tichai2022combining,stetcu,Gu:2023aoc} or the quantum magic~\cite{brokemeier2024} .
A remarkable finding associated to this line of research is the recent finding that, among all possible bipartitions, the separation of a nuclear model space into proton and neutron orbitals has the lowest von Neumann entropy~\cite{perezobiol2023}. Morevoer, proton-neutron entanglement decreases for more neutron-rich systems~\cite{johnson2023proton}. These properties have been used to improve nuclear-structure calculations using classical methods~\cite{Tichai:2024cyd,Gorton:2024hbb}.

\begin{figure}[t]
     \centering
     \includegraphics[width=0.96\linewidth]{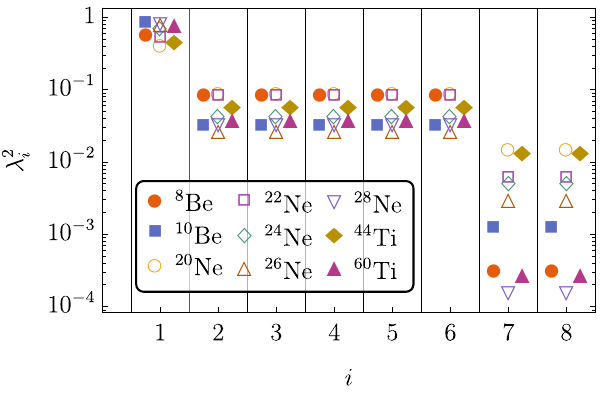}
     \caption{First eight singular values, $\lambda_i$, of the Schmidt decomposition into protons and neutrons for the NSM ground state of different beryllium (\emph{p} shell, solid circles and squares), neon (\emph{sd} shell, empty symbols) and titanium (\emph{pf} shell, solid diamond and triangle) nuclei. In all cases, a first large singular value is followed by much smaller, five-fold degenerate second to seventh singular values. The Schmidt coefficients for these degenerate singular values are smaller in neutron-rich isotopes.
     } 
     \label{fig:svs}     
\end{figure}

Here, we apply these insights into the Schmidt decomposition of Eq.~(\ref{eq:schmidt}) with the aim of improving quantum simulations of nuclear structure. Figure~\ref{fig:svs} shows the first eight singular values for various beryllium~\footnote{We calculate beryllium isotopes employing the Cohen-Kurath interaction~\cite{cohen1965effective} in the $p-$shell. We do not shown the equivalent to Fig.~\ref{fig:shells} in this configuration space for brevity.}, neon, and titanium isotopes. In all nuclei, the singular values become smaller exponentially, especially for the most neutron-rich isotopes ${}^{28}$Ne and ${}^{60}$Ti. The Schmidt decomposition consists of a first large non-degenerate value, with $\lambda_1$ close to unity, followed by five degenerate vectors with coefficients $10^{-2}<\lambda_{2-6}^2<10^{-1}$. This picture is qualitatively similar to that found in the FH model in Fig.~\ref{fig:svs_FH}, except that the second set of states has $5$ rather 
than $4$ degenerate states. 

The degeneracy in the NSM can be associated with five product states with opposite $M$ in the proton and neutron states, which we denote $|J_z^{(p)}\rangle\otimes|J_z^{(n)}\rangle$: $|\pm 2\rangle\otimes|\mp 2\rangle$, $|\pm 1\rangle\otimes|\mp 1\rangle$, and $|0\rangle\otimes|0\rangle$. These results suggests that nuclear ground states, especially for neutron-rich isotopes, can be well approximated by six separate product states of only protons and neutrons. As a consequence, and due to this degeneracy and the overall normalization of the state, a single coefficient is needed to describe the full state in terms of very few Schmidt vectors. 

The Schmidt coefficients beyond $i>6$ in Fig.~\ref{fig:svs} are relatively small compared to the first $6$ eigenvalues. For the Ne and Ti isotopes shown in the figure, we find that as the systems become more  neutron rich, the corresponding Schmidt coefficients for $i>6$ become smaller. Whereas $\lambda_7^2 \approx 10^{-2}$ for $^{20}$Ne, the same coefficient for ${}^{28}$Ne is two orders of magnitude smaller, $\lambda_7^2\approx 10^{-4}$. A similar pattern is observed in titanium, between ${}^{44}$Ti and ${}^{60}$Ti. In the following, we choose ${}^{28}$Ne and ${}^{60}$Ti as our reference isotopes for the EDEF analysis, since we expect these to be best suited for the approach. 

In addition, nuclear entanglement can also be partly understood in terms of subshell closures and occupation numbers. Nuclei with proton or neutron orbitals that are mostly empty or fully-occupied present very low entanglement between these orbitals and the rest of states~\cite{perezobiol2023}. This feature motivates us to apply the EDEF approach in an iterative fashion. Firstly, we shall use a proton-neutron partition. Second, we employ an additional subpartition based on single-particle energies, separating the higher- and lower-energy  states within each proton and neutron subsystem. Figure~\ref{fig:shells} illustrates the two partitions and the corresponding cuts used for ${}^{28}$Ne and ${}^{60}$Ti, which are the two isotopes of reference discussed in the following Section. For the latter nucleus, a bipartition into subsystems of the same size requires separating the energy-degenerate single-particle states of the $1p_{3/2}$ orbital.

\section{Entropy-driven entanglement forging with ADAPT-VQE}\label{sec:quantum_algorithm}

ADAPT-VQE~\cite{grimsley2019adaptive,tang2021qubit,anastasiou2024tetris,ramoa2024reducing,Feniou2023} is a variational quantum algorithm which updates iteratively a user-defined ansatz, rather than optimizing a fixed number of parameters as the UCC-VQE~\cite{stetcu,papenbrock,uccrev}. Each iteration $k$ adds a new unitary operator, $A_k = e^{i \theta_k T_k}$, to the ansatz, with a new parameter, $\theta_k$, and an Hermitian operator $T_k$ from a predefined operator pool. Then, all parameters, $\bm{\theta}=\{\theta_1,\cdots,\theta_k\}$, are optimized simultaneously to minimize the energy,
\begin{equation}
    E_{\textrm{ADAPT-VQE}} = \min_{\bm{\theta}} \frac{\langle \psi(\bm{\theta}) | H_{\rm{eff}} |\psi(\bm{\theta})\rangle}{\langle \psi(\bm{\theta}) |\psi(\bm{\theta})\rangle}.
\end{equation}
For iteration $k$, the optimization starts with the parameters values $\theta_1,\cdots,\theta_{k-1}$ obtained in the minimization of the previous iteration, $k-1$, and with the new parameter set to zero, $\theta_k=0$. 
The operator $T_k$ is chosen such that the energy gradient with respect to $\theta_k$,
\begin{equation}
    \left. \frac{\partial E^{(n)}}{\partial \theta_k } \right\vert_{\theta_k=0} = 
    \left. i \langle \psi (\bm{\theta})| [H_{\rm{eff}},A_k] |\psi (\bm{\theta})\rangle \right\vert_{\theta_k=0}\,, 
    \label{eq:gradient}
\end{equation}
is maximum. Thus, by starting from the minimum in the $(k-1)$-dimensional parameter space and maximizing the gradient in the new dimension, the algorithm gradually increases the parameter space while always decreasing the energy. ADAPT-VQE was originally proposed to solve the electronic structure of the ground state of molecules~\cite{yordanov2021qubit,grimsley2023adaptive}, and since then it has been applied to a broad variety of quantum many-body systems, including  lattice quantum electrodynamics~\cite{illa2,illa3}, the NSM~\cite{romeroquantum,PerezMarquez2023}, algebraic nuclear models~\cite{romeroquantum,baid2024extended} and nuclear pairing~\cite{Zhang:2024uxp}. In addition to the novel FH simulations, we build on our previous work~\cite{PerezMarquez2023,perezobiol2023} and optimize the ADAPT-VQE  algorithm to solve NSM ground states with fewer quantum resources by exploiting EDEF.

ADAPT-VQE requires a choice of reference state and a pool of operators.
For the latter, we take all the two-body hopping operators,
\begin{equation}\label{eq:ferm_pool_2b}
    T_{rs}^{pq} = i (a_p^{\dag} a_q^{\dag} a_r a_s - a_r^{\dag} a_s^{\dag} a_p a_q),
\end{equation}
with the restriction that they conserve the  Hamiltonian symmetries: the spin for the FH and $J_z,T_z$ for the NSM. For the FH, in addition, we also include one-body operators,
\begin{equation}\label{eq:ferm_pool_1b}
    T_{s}^{r} = i (a_r^{\dag} a_s - a_s^{\dag}  a_r),
\end{equation}
to the operator pool. 

\begin{figure*}[t]
     \centering
     \includegraphics[width=1\linewidth]{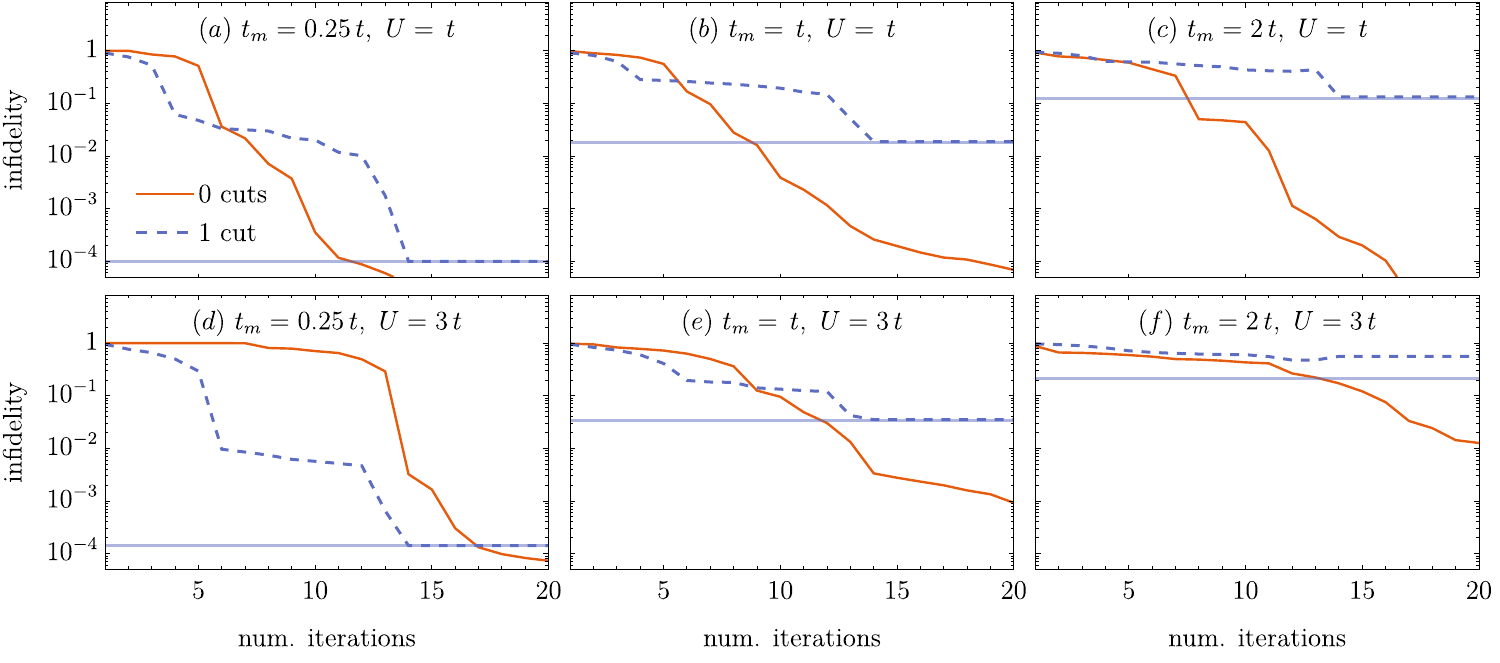}
     \caption{Infidelity as a function of the number of ADAPT-VQE iterations for the same FH model parameters (middle hopping $t_m$, interaction $U$) as in Fig.~\ref{fig:svs_FH}. ADAPT-VQE without entanglement forging (solid red lines) is compared with one layer of EDEF (labelled ``1 cut'', dashed blue lines). Horizontal lines  
     indicate the infidelity minimum set by the Schmidt-decomposition cutoff. 
     } 
     \label{fig:results_FH}     
\end{figure*}

As a reference state, with the standard ADAPT-VQE we choose the lowest-energy Slater determinant in the Fock basis of the particular FH lattice or nucleus. This can be implemented with one-qubit gates under the Jordan-Wigner mapping~\cite{perez2022digital}.  
With EDEF, however, we can exploit other physical arguments to choose a more convenient reference state, such as degeneracies in the Schmidt decomposition that strongly suggest specific symmetries. In the case of the FH model, for instance, the best choice is a symmetric distribution of up and down spins between the left ($l$) and right ($r$) partitions of the lattice,
\begin{align}
    |\psi\rangle=&
    \lambda_0|l_{\uparrow\downarrow}\rangle\otimes|r_{\uparrow\downarrow}\rangle
  + \lambda_{1}\bigg( 
|l_{\uparrow\uparrow\downarrow}\rangle\otimes|r_{\downarrow}\rangle
  + |l_{\downarrow}\rangle\otimes|r_{\uparrow\uparrow\downarrow}\rangle \nonumber
  \\&
  + |l_{\uparrow\downarrow\downarrow}\rangle\otimes|r_{\uparrow}\rangle
  + |l_{\uparrow}\rangle\otimes|r_{\uparrow\downarrow\downarrow}\rangle
\bigg).
\label{eq:schmidt5}
\end{align}
This particular choice is inspired by the Schmidt decomposition of the full state presented in Fig.~\ref{fig:svs_FH}. The analysis of the associated
quantum numbers of each Schmidt vector indicates that this choice should provide an excellent reconstruction of the full state as measured
by the infidelities of Fig.~\ref{fig:fh_infs}. 

In turn, for the NSM we use eigenstates of the $J_z$ operator in the proton ($p_{J_z}$) and neutron ($n_{J_z}$) sectors. The pattern for the Schmidt decomposition described in Sec.~\ref{sec:nsm} and shown in Fig.~\ref{fig:svs} suggests the following choice of reference state,
\begin{align}
    |\psi\rangle=&
    \lambda_0|\tilde{p}_0\rangle\otimes|\tilde{n}_0\rangle
  + \lambda_{1}\bigg( 
    |p_{-2}\rangle\otimes|n_{2}\rangle
  + |p_{-1}\rangle\otimes|n_{1}\rangle \nonumber
  \\&
  + |p_{0}\rangle\otimes|n_{0}\rangle
  + |p_{1}\rangle\otimes|n_{-1}\rangle
  + |p_{2}\rangle\otimes|n_{-2}\rangle\bigg).
  \label{eq:schmidt6}
\end{align}
We note that for both the FH and the NSM applications, we employ reference states that only involve two coefficients, $\lambda_0$ and $\lambda_1$, which we fix to the corresponding values of the Schmidt decomposition (see Appendix~\ref{app:layers} for more details).

In order to ensure orthogonality in the NSM for the states with the same ${J_z}^{(p)}$ and ${J_z}^{(n)}$, we start from two orthogonal Slater determinants and use the same choice of operators for both states. Since the operator pool preserves the quantum numbers for each subsystem, orthogonality will be maintained through the iterations. The second layer of EDEF splits the proton (neutron) nuclear ground state into symmetric distributions of protons (neutrons) and proton holes (neutron holes). Appendix \ref{app:layers} gives further details on the properties of these states and their corresponding coefficients, illustrated with the specific examples used in our optimization. Having defined the operator pools and reference states of our EDEF-accelerated ADAPT-VQE, we proceed to discuss the results of this approach for both the FH and the NSM. 

\section{Results}\label{results}
\subsection{Fermi-Hubbard model}
\label{sec:fh_res}

In the FH model, we can access systems with a wide range of entanglement across left and right partitions by tuning the parameter $t_m$. This allows us to analyze the quality of the EDEF method depending on the entanglement structure of the system. For this purpose, like in Sec.~\ref{sec:fh} and Fig.~\ref{fig:svs_FH}, we simulate systems with three different central hopping values, $t_m=0.25t,t,2t$, and two interactions, $U=t,3t$. We obtain the ground state of these systems using two methods. First, we employ the full, regular ADAPT-VQE to solve for the system ground state. Second, we use an optimization based on EDEF, as described in Sec.~\ref{sec:quantum_algorithm}. 
To compare the efficiency of the two approaches, we focus on the infidelities of each approach with respect to the exact ground state,
\begin{equation}
I= 1-|\langle\psi_{\xi}|\psi_{\text{exact}}\rangle|^2 \, ,
\end{equation}
where $\xi$ can correspond to either the full ADAPT-VQE approach or to an approximate EDEF with $1$ or $2$ cuts. 
An alternative analysis using the ground-state energy relative error,
\begin{equation}
\epsilon_E= \left|\frac{E_{\xi}-E_{\text{exact}}}{E_{\text{exact}}}\right|,
\end{equation}
gives very similar results.

Figure~\ref{fig:results_FH} compares the infidelities obtained with the regular ADAPT-VQE without entanglement forging (solid red lines) and with EDEF with one layer, or ``1-cut'' (dashed blue lines). This approach corresponds to a reference state involving five singular values, as described in Sec.~\ref{sec:fh}.
The central hopping strength increases from left to right panels. Top panels show results for $U=t$, and bottom panels correspond to a stronger interaction $U=3t$. The infidelity in logarithmic scale is shown as a function of the iteration number for both ADAPT-VQE and EDEF. In practical terms, the ADAPT-VQE simulation corresponds to a single circuit and each iteration $k$ involves the addition of a new operator $A_k$ and the optimization of $k$ parameters, $\theta_1, \cdots, \theta_k$. 
We define one iteration in the 1-cut EDEF to the addition of a single new operator in either of the different subcircuits which we employ to describe the full system (see Appendix~\ref{app:layers} for details on the nature and number of subcircuits). 
Horizontal lines mark the lower bound for our ansatz infidelities. This is given by the infidelity associated to the corresponding Schmidt decomposition of the ground state with cutoff at the fifth product state, $\chi_{\rm cut}=5$ in Eq.~(\ref{eq:schmidt5}). With the degeneracy, only two different Schmidt coefficients are involved. 

Even though there is no variational bound on the decrease of infidelities as there is for the energy, we find that both ADAPT-VQE and 1-cut EDEF results reduce the infidelity as the number of iterations increase. Regular ADAPT-VQE converges to $I\lesssim 1 \%$ within $20$ iterations for all cases. The convergence is however relatively slower for strongly-interacting systems ($U=3t$, bottom panels) compared to less correlated systems ($U=t$, top panels). In contrast, the interaction strength $U$ makes no difference for the 1-cut EDEF results. For the more physical case of $t_m=t$, for instance, the 1-cut EDEF results converge to the lower bound within $14$ iterations. We stress that almost all 1-cut EDEF simulations converge to this lower bound, which highlights the good performance of the variational optimization. The exception corresponds to the extreme case of $t_m=2t$ and $U=3t$ (lowest right panel), which has the largest interaction values across all panels. In this case, the ansatz has a hard time converging to the exact wavefunction even for the full ADAPT-VQE simulation. 

\begin{table}[t]
\begin{tabular}{c|c|c|c|c|c|c}
                             & { cuts } & { $N_q$ } & $~N_\text{it}~$ &  $\epsilon_E$  & $I_\text{conv}$                  & $r$     \\ \hline
\multirow{2}{*}{(0.25,1)}    & 0    & 8            & 16  & $7.4\times10^{-6}$       & $5.6\times10^{-6}$ & 0.74  \\ 
                             & 1    & 4            & 14  & $1.1\times10^{-4}$       & $9.9\times10^{-5}$ & 0.61  \\ \hline
\multirow{2}{*}{(1,1)}       & 0    & 8            & 24  & $1.4\times10^{-5}$       & $8.0\times10^{-6}$ & 0.47  \\ 
                             & 1    & 4            & 14  & $1.9\times10^{-2}$       & $1.9\times10^{-2}$ & 0.26  \\ \hline
\multirow{2}{*}{(2,1)}       & 0    & 8            & 19  & $1.6\times10^{-5}$       & $9.5\times10^{-6}$ & 0.58  \\ 
                             & 1    & 4            & 14  & $1.2\times10^{-1}$       & $1.3\times10^{-1}$ & 0.13  \\ \hline
\multirow{2}{*}{(0.25,3)}    & 0    & 8            & 24  & $1.8\times10^{-5}$       & $9.4\times10^{-6}$ & 0.46  \\ 
                             & 1    & 4            & 14  & $1.5\times10^{-4}$       & $1.4\times10^{-4}$ & 0.59  \\ \hline
\multirow{2}{*}{(1,3)}       & 0    & 8            & 31  & $2.0\times10^{-5}$       & $5.4\times10^{-6}$ & 0.35  \\ 
                             & 1    & 4            & 14  & $3.2\times10^{-2}$       & $3.5\times10^{-2}$ & 0.22  \\ \hline
\multirow{2}{*}{(2,3)}       & 0    & 8            & 34  & $3.2\times10^{-6}$       & $3.7\times10^{-6}$ & 0.37  \\ 
                             & 1    & 4            & 14  & $3.8\times10^{-1}$       & $5.6\times10^{-1}$ & 0.039 \\ \hline \hline
\multirow{3}{*}{${}^{28}$Ne} & 0    & 24           & 100  & $6.2\times10^{-3}$      & $1.0\times10^{-1}$ & 0.023 \\ 
                             & 1    & 12           & 85  & $6.0\times10^{-4}$       & $2.9\times10^{-3}$ & 0.069 \\ 
                             & 2    & 6            & 48  & $9.8\times10^{-3}$       & $5.1\times10^{-2}$ & 0.062 \\ \hline
\multirow{2}{*}{${}^{60}$Ti} & 1    & 12           & 57  & $1.7\times10^{-1}$       & $2.5\times10^{-1}$ & 0.024 \\ 
                             & 2    & 6            & 42  & $1.6\times10^{-1}$       & $8.2\times10^{-2}$ & 0.059 \\
\end{tabular}
\caption{Number of qubits ($N_q$) per circuit used to simulate the system with regular (0 cuts) and EDEF optimized (1 or 2 cuts) ADAPT-VQE, as well as the number of iterations ($N_\text{it}$),  relative error in the energy ($\epsilon_E$), infidelity ($I_\text{conv}$) and convergence rate ($r$) once the algorithm has either converged ($I_{\rm conv}<10^{-5}$ for ADAPT-VQE) or reached the maximum number of iterations imposed. Top: Results for FH lattices labeled as ($t_m$, $U$). Bottom: Results for the NSM simulations of ${}^{28}$Ne and ${}^{60}$Ti. }
\label{tab:results}
\end{table}

The left panels of Fig.~\ref{fig:results_FH} show that for weakly linked FH lattices, $t_m=0.25t$, 
one layer of EDEF offers a clear advantage for $U=3t$ and target infidelities $\epsilon_E\gtrsim10^{-4}$. For a weaker interaction $U=t$, the 1-cut EDEF infidelities also improve over regular ADAPT-VQE up to $I\sim 10^{-2}$ (or seven iterations). Beyond this point, ADAPT-VQE converges faster.
For the canonical FH lattice, $t_m=t$ (central panels), the entropy is still low -- see Fig.~\ref{fig:fh_infs} -- and Fig.~\ref{fig:results_FH} indicates that 1-cut EDEF reaches almost $I\sim10^{-2}$, although converging more slowly than the regular ADAPT-VQE. Figure~\ref{fig:results_FH} also shows that, when $t_m>t$ (right panels), EDEF starts to underperform substantially compared to ADAPT-VQE. This can be associated to a larger entanglement entropy. For these systems, reaching $I\lesssim 10^{-3}$ may require to train more than eight copies of the circuit which, at this system size, does not provide a clear performance advantage over using all the qubits like in the regular ADAPT-VQE -- although the local optimization could still provide some benefits. We postpone a thorough analysis of these issues for further work.

Table~\ref{tab:results} (top rows) quantifies the resources and performance of the FH simulations using regular ADAPT-VQE and the EDEF optimization. It lists the number of iterations, $N_\text{it}$, the infidelity, $I_\text{conv}$, and energy relative error, $\epsilon_E$, once the optimization has converged. As a convergence criterion, we stop our ADAPT-VQE simulations when the infidelity falls below $I < 10^{-5}$. For EDEF, convergence is often found above this threshold value, when gradients reach very small values and updates do not improve the energy any further. In other cases, particularly for the more demanding NSM simulations, we stop the simulations arbitrarily after $100$ iterations. 

Table~\ref{tab:results} presents a proxy for the state of the convergence procedure, the so-called convergence rate, which we define as
\begin{equation}
r\equiv\frac{-\log(I_\text{conv})}{N_\text{it}}\,.
\label{eq:rate}
\end{equation}
Consistently with Fig.~\ref{fig:results_FH}, the 1-cut EDEF convergence rates are higher for weaker central hopping values, but in general they are lower than for the regular ADAPT-VQE. Nonetheless, Table~\ref{tab:results} also highlights that all 1-cut simulations require $4$ qubits and converge with just $14$ iterations, while all regular ADAPT-VQE optimizations need all $8$ qubits and $16$ or more iterations. In contrast, the infidelity of the converged results reaches a high quality for ADAPT-VQE, with $I\leq10^{-5}$ across different values of $t_m$ and $U$. The best EDEF simulations reach at most $I \approx 10^{-4}$, and often reach worse values. Having said that, circuits with half as many qubits also constrain fermion operators to be more local. Therefore, in addition to reducing the amount of qubits needed, these require fewer quantum gates, which is another advantage of the EDEF implementation, as we analyse in Sec.~\ref{sec:gates}.

\subsection{Nuclear shell model}\label{sec:nuclear_res}

We now apply the EDEF method to study neutron-rich isotopes ${}^{28}$Ne and ${}^{60}$Ti within the NSM. These are ideal systems for EDEF, because of the low entanglement between the neutron and proton partitions, as shown in Fig.~\ref{fig:svs}. Moreover, these are challenging isotopes from the regular ADAPT-VQE perspective, because they involve a relatively large number of nucleons and single-particle states. In fact, in our previous work~\cite{PerezMarquez2023} we were not able to simulate ${}^{60}$Ti with ADAPT-VQE because of the large memory required to encode the corresponding state vectors. 

We explore the performance of EDEF using various layers. In particular, we use ADAPT-VQE with circuits with $N_q$ (regular ADAPT-VQE), $N_q/2$ (EDEF with one cut), and $N_q/4$ (EDEF with two cuts) qubits. $N_q$ is the number of single-particle states in the valence space, which in the mapping described in Fig.~\ref{fig:shells} also corresponds to the number of qubits. In the $sd$ shell, we require $N_q=24$ qubits to simulate ${}^{28}$Ne, whereas for ${}^{60}$Ti in the $pf$ shell, $N_q=40$ qubits are needed. Similarly to Sec.~\ref{sec:fh_res}, we quantify the quality of both the full ADAPT-VQE and the EDEF approach in terms of the infidelity with respect to the exact ground state obtained with each approach. 

\begin{figure}[t]
     \centering
     \includegraphics[width=0.95\linewidth]{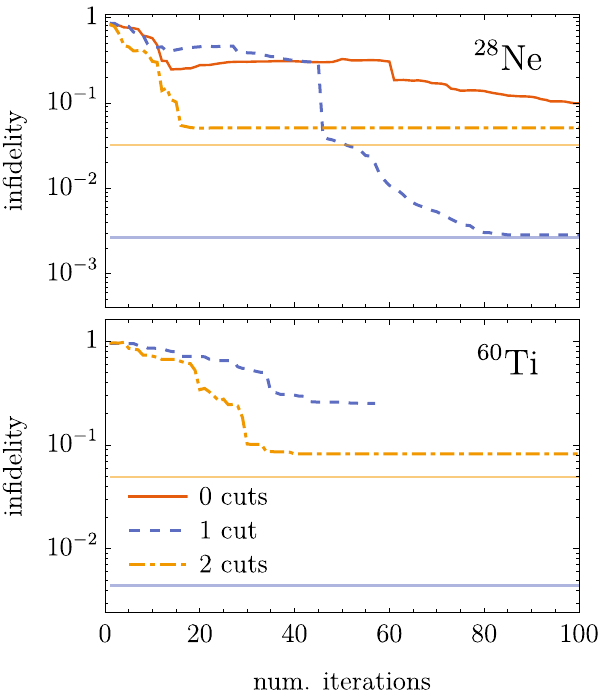}
     \caption{Infidelities as a function of the number of ADAPT-VQE iterations for NSM simulations of ${}^{28}$Ne (top panel) and ${}^{60}$Ti (bottom panel), using the regular ADAPT-VQE (0 cuts, solid red line) and the optimized EDEF with one (1 cut, dashed blue lines) or two (2 cuts, dashed-dotted yellow lines) cuts. Horizontal lines with the same colour code indicate the infidelities set by the Schmidt decompositions cut off consistently with each EDEF simulation. 
     } 
     \label{fig:netierrors}     
\end{figure}

The top panel of Fig.~\ref{fig:netierrors} presents the evolution of the infidelity of the ground-state wave function of  ${}^{28}$Ne as a function of the number of ADAPT-VQE iterations for the regular ADAPT-VQE (solid red line), 1-cut EDEF (dashed blue line) and 2-cut EDEF (dashed-dotted yellow line). As for the FH model, the infidelities for one and two layers of EDEF converge very close to the lower bounds defined by the Schmidt decomposition (horizontal lines of the corresponding color). The two EDEF variants converge faster than the regular ADAPT-VQE, meaning that with the same number of parameters, and half or one quarter of the qubits per circuit, EDEF outperforms ADAPT-VQE. We note that the full ADAPT-VQE simulations increase the infidelity with respect to the exact ground state between iterations $15$ and $60$, where a sudden drop occurs and the decline of infidelity resumes. This is possible because ADAPT-VQE is a variational optimisation process based on the minimization of energies, so the infidelity can increase when not too close to the target energy -- unless the ground state space is degenerate or the energy spectrum is gapless, the infidelity must eventually decrease as the error in energy approaches zero. The 1-cut EDEF shows a much milder increase of infidelity between iterations $15$ and $25$. It also converges to much lower infidelities, $I \approx 10^{-3}$, than the overall 2-cut EDEF simulation. This happens, however, after a significantly larger number of iterations. Indeed, the results in Table~\ref{tab:results} indicate that the convergence rate is similar in both cases.   

The bottom panel of Fig.~\ref{fig:netierrors} shows the results for ${}^{60}$Ti with one and two layers of EDEF, with the same line and color code as the top panel. For this nucleus, standard ADAPT-VQE with no cuts exceeds our computational capabilities and is thus not shown here.
The infidelity for 1-cut EDEF shows good convergence up to $\sim50$ iterations. 
At this point, when $I$ is still well above the limit set by the Schmidt decomposition, the computation becomes too slow for our current computational resources and we stop it.  
In contrast, the 2-cut EDEF requires significantly less resources. We find that this 2-cut EDEF simulation is able to converge to a high-quality local minimum at the 42$^{\rm{nd}}$ iteration, with an infidelity of $I=0.082$. The corresponding relative error in the energy is $\epsilon_E=0.16$. Consistently, Table~\ref{tab:results} indicates a better convergence in terms of infidelities, energies and rates for the 2-cut simulation as opposed to the 1-cut approach for this isotope.

In principle, nothing precludes the application of EDEF to less neutron-rich nuclei. However, in general we expect these to have larger proton-neutron entanglement and thus the EDEF approach to work less efficiently. For instance, Fig.~\ref{fig:svs} shows that ${}^{26}$Ne has a more complex structure in terms of product states than ${}^{28}$Ne, with significantly larger $i=7$ and $8$ Schmidt  coefficients. Nonetheless, the one-layer EDEF achieves only marginally poorer infidelity and relative energy error than the regular ADAPT-VQE, but using half as many qubits and fewer resources. We provide the ADAPT-VQE and EDEF results for ${}^{26}$Ne in Appendix~\ref{app:26ne}.

\subsection{Number of CNOT gates in EDEF NSM simulations}\label{sec:gates}
An important factor for the implementation of the current algorithm in quantum computers is the depth of the circuit, which is largely dependent on the number of two-qubit gates needed. This is specially relevant for NSM simulations, which already for medium-mass nuclei require a very significant amount of resources well beyond current capabilities~\cite{PerezMarquez2023,Li:2023eyg}. 

\begin{figure}[t]
     \centering
     \includegraphics[width=1\linewidth]{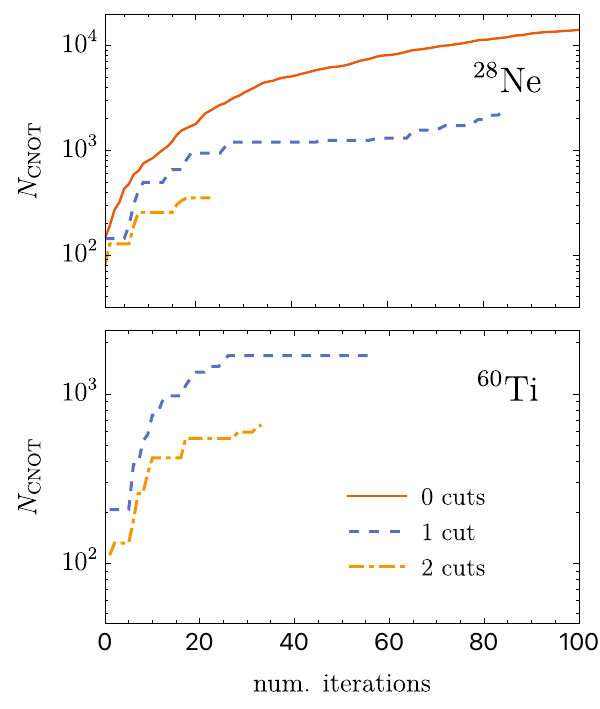}
     \caption{Number of CNOT gates for the circuit with maximum number of them for the simulation of ${}^{28}$Ne (top) and  ${}^{60}$Ti (bottom) as a function of the ADAPT-VQE iterations. The results compare the regular ADAPT-VQE (0 cuts, solid red line) with EDEF with one layer (1 cut, blue dashed line) and two layers (2 cuts, dashed-dotted yellow lines).
     } 
     \label{fig:28necx}     
\end{figure}

Figure~\ref{fig:28necx} compares the number of CNOT gates, $N_{\rm CNOT}$, required in the NSM simulation of ${}^{28}$Ne and ${}^{60}$Ti in the case of regular ADAPT-VQE and EDEF with one or two layers. Since EDEF involves more than one circuit, we select the circuit with the maximum number of CNOT gates to plot Fig.~\ref{fig:28necx}. The top panel shows that, for ${}^{28}$Ne, the number of CNOTs in the $24$-qubit circuit (ADAPT-VQE, solid red line) increases drastically with the number of iterations. By the $100^{\rm th}$ iteration, more than $10^4$ CNOT gates are required. This number is however drastically reduced by one order of magnitude throughout the whole optimization when we employ EDEF with one layer, using $12$ qubits (1-cut EDEF, dashed blue line). Moreover, the $6$-qubit EDEF with two layers (2-cut EDEF, dotted-dashed yellow line) reduces the number of CNOTs by an additional order of magnitude. Consistently, the bottom panel of Fig.~\ref{fig:28necx} indicates that for ${}^{60}$Ti the number of CNOTs is smaller in the 2-cut EDEF, which uses $10$-qubit circuits, than in the 1-cut EDEF, involving $20$ qubits.
These results highlight that our EDEF approach, in addition to having better performance than the regular ADAPT-VQE as shown in Sec.~\ref{sec:nuclear_res}, also requires substantially fewer quantum resources for these systems. As a consequence, the method is specially suited for near-future quantum device implementations. 

The reasons underlying the significant reduction of CNOT gates within the EDEF approach are twofold. First, each EDEF cut makes circuits considerably shallower because at each iteration an operator is added to a single circuit. This distributes the CNOT gates among all circuits, rather than accumulating them in a single one. Second, circuits with fewer qubits constrain the fermionic operators in the ADAPT-VQE pool. Because of the cuts, these operators are more local ({\it eg} connecting nearby qubits). This requires less CNOT gates in the corresponding Jordan-Wigner mapping, reducing the overall costs.

Among all the EDEF subcircuits describing both ${}^{60}$Ti and ${}^{28}$Ne, the ones with most CNOT gates are those simulating the first, non-degenerate product state in the Schmidt decomposition, $\lambda_0|\tilde{p}_0\rangle\otimes|\tilde{n}_0\rangle$ in Eq.~(\ref{eq:schmidt6}). 
This is because the ADAPT-VQE algorithm finds, in most iterations, largest gradients for operators added to the first $\langle J_z \rangle =0$ proton-neutron product state. This is expected from the Schmidt decomposition, as the first singular value is much larger than the rest and should therefore
 contribute most to lowering the ground-state energy, at least in the initial iterations. 
Moreover, we also find that ADAPT-VQE chooses the same circuit for many contiguous iterations, showing plateaus that indicate a constant number of CNOTs in all but one circuit during various iterations. We provide more details on the specifics of the ADAPT-VQE optimization with EDEF in Appendix~\ref{app:optimization}, where we also 
break down the number of CNOT gates for each of the circuits that appear in the EDEF approach.

\section{Summary and outlook}\label{conclusions}

We present a novel procedure, entropy-driven entanglement forging (EDEF), that optimizes the solution of many-body problems in quantum computers using VQAs. Our approach is based on entanglement forging, which reduces the number of qubits and quantum gates needed to solve the problem. Crucially, in order to decide on the corresponding partitions and coefficients, we advocate for physical criteria associated to the entropy between partitions of the system. To this end, we explore systems that we expect to have low entropy partitions and employ the Schmidt decomposition as a tool to characterize them through their different quantum numbers. We build reference states that are linear combinations of smaller circuits with the quantum numbers associated to the Schmidt decomposition. Moreover, we exploit degeneracy in the lowest-lying Schmidt vectors to employ a very small number of subcircuits and coefficients, requiring fewer and more local resources. 

With this technique, we have successfully simulated ground states of FH lattices with various central hoppings and interactions, and of the isotopes ${}^{28}$Ne and ${}^{60}$Ti calculated with the NSM. In both cases, we compare optimizations using ADAPT-VQE with and without EDEF. For both types of systems, EDEF exploits the low entanglement between different partitions of the system. 
In the FH model, we can tune the entanglement between left and right parts of the lattice through a central hopping term with variable strength, displaying the power of the technique on different settings. 
In contrast, we exploit the recently observed small entanglement  between protons and neutrons in the NSM to improve upon previous simulation attempts. This knowledge is combined with the singular-value degeneracy in the Schmidt decomposition of the ground states. Our EDEF simulations with one layer of EDEF -- separating the system in two unique partitions -- only need half as many qubits as the regular ADAPT-VQE, and they describe nuclei with an order of magnitude fewer CNOT gates. Nuclei simulated with two EDEF layers, with each bipartition additionally separated into two parts corresponding to higher- and lower-energy single-particle states, need a quarter of the regular ADAPT-VQE qubits and an additional order of magnitude fewer CNOTs.

In terms of performance, in FH lattices, we find that EDEF converges better than regular ADAPT-VQE for weak central hopping. The approach is also more efficient in terms of iterations as long as the central hopping is close to the other hoppings, $t_m\approx t$. In the NSM, for ${}^{28}$Ne, one layer of EDEF using 8 circuits is more efficient than the regular ADAPT-VQE. 
Furthermore, EDEF allows us to simulate ${}^{60}$Ti, which is beyond our capabilities with the standard ADAPT-VQE. For this nucleus, the two layer EDEF reaches a lower infidelity than the 1-cut EDEF.

In summary, EDEF is an approach ideally suited to accommodate the qubit and CNOT gate number limitations present in current intermediate-scale noisy devices. In particular, EDEF allows one to adjust algorithms to produce shallower circuits, mitigating the impact of errors. This reduction can be further improved by the adaptability of the trained parameters expected for a variational algorithm. Furthermore, by decreasing the number of qubits, EDEF allows one to use smaller devices with lower error rates. In principle, we have shown that these advantages can be exploited further by identifying low-entanglement partitions in successive EDEF layers. 
Beyond the first applications studied in this work, EDEF can be used to study other many-body systems for which low-entropy subsystems are identified. More generally, EDEF can potentially be used to optimize other types of VQAs for different applications, including for instance quantum machine learning settings. 

\acknowledgments

This work is financially supported by 
MCIN/AEI/10.13039/501100011033 from the following grants: PID2020-118758GB-I00, PID2021-127890NB-I00, PID2023-147112NB-C22, RYC-2017-22781 and RYC2018-026072 through the “Ram\'on
y Cajal” program funded by FSE “El FSE invierte en tu futuro”, CNS2022-135529 and CNS2022-135716 funded by the
“European Union NextGenerationEU/PRTR”, and CEX2019-000918-M to the “Unit of Excellence Mar\'ia de Maeztu 2020-2023” award to the Institute of Cosmos Sciences; and by the Generalitat de Catalunya, grants 2021SGR01095 and 2021SGR00907. 

\bibliography{biblio}

\appendix

\section{Initial states and coefficients of EDEF}\label{app:layers}
In the main text we outlined the initial states needed for EDEF. In this section we give extended detail and a few examples for both the first layer and second layer of EDEF.

\subsection{First layer of entanglement forging}
For a FH lattice of four sites, we use five pairs of circuits with four qubits, instead of a single circuit with eight qubits. Each circuit simulates a quantum state term in the product of the truncated (and renormalized) Schmidt decomposition:
\begin{equation}\label{eq:schmidt5A}
\begin{aligned}
    |\psi\rangle=&
    \lambda_0|l_{\uparrow\downarrow}\rangle\otimes|r_{\uparrow\downarrow}\rangle
  + \lambda_{1}\big( 
|l_{\uparrow\uparrow\downarrow}\rangle\otimes|r_{\downarrow}\rangle
  + |l_{\downarrow}\rangle\otimes|r_{\uparrow\uparrow\downarrow}\rangle
  \\&
  + |l_{\uparrow\downarrow\downarrow}\rangle\otimes|r_{\uparrow}\rangle
  + |l_{\uparrow}\rangle\otimes|r_{\uparrow\downarrow\downarrow}\rangle
\big).
\end{aligned}
\end{equation}
A single circuit of four qubits, for example, simulates the state $|l_{\uparrow\downarrow}\rangle$ in the first product state, which simulates the left part of the lattice, labeled with $l$, and which has one spin up fermion and one with spin down, labeled in the subindex as $\uparrow\downarrow$. Note that all states are naturally orthogonal as they involve different distributions of particles and spins. This orthogonality is maintained as operators from the pool, which conserve spin and particle number, are added to each circuit.

For the nuclear shell model and for all isotopes considered, with one layer of entanglement forging we separate the proton and neutrons degrees of freedom as shown in Fig.~\ref{fig:shells} and thus reduce by half the number of qubits of the simulating circuits. As observed in Fig.~\ref{fig:svs}, the nuclear state can be very accurately approximated using the following truncation in the Schmidt decomposition
\begin{equation}\label{eq:schmidt6A}
\begin{aligned}
    |\psi\rangle=&
    \lambda_0|\tilde{p}_0\rangle\otimes|\tilde{n}_0\rangle
  + \lambda_{1}\big( 
    |p_{-2}\rangle\otimes|n_{2}\rangle
  + |p_{-1}\rangle\otimes|n_{1}\rangle
  \\&
  + |p_{0}\rangle\otimes|n_{0}\rangle
  + |p_{1}\rangle\otimes|n_{-1}\rangle
  + |p_{2}\rangle\otimes|n_{-2}\rangle\big).
\end{aligned}
\end{equation}
With one circuit, for example, we simulate the state 
$|p_{-2}\rangle$, which corresponds to a superposition
of Slater determinants containing only proton orbitals
and with expected third component of total angular momentum $\langle p_{-2}|J_z|p_{-2}\rangle=-2$.
We need to ensure these states are orthogonal. All states that have different $\langle J_z \rangle$ in each partition are already orthogonal, and will continue to be considering the operators in our pool. This is however not the case for the two states with $\langle J_z \rangle=0$,  $|\tilde{p}_0\rangle\otimes|\tilde{n}_0\rangle$ and $|p_0\rangle\otimes|n_0\rangle$.
For these, we start with different (orthogonal) Slater determinants, $|\psi_0\rangle$, $|\tilde{\psi}_0\rangle$, and then apply the same unitaries to both states to keep them orthogonal, that is, 
$\langle\tilde{\psi}_0| e^{-i\theta T} e^{i\theta T} |\psi_0\rangle=
\langle\tilde{\psi}_0 |\psi_0\rangle=0$.

At each iteration the gradients are computed for each of the terms in Eqs.~(\ref{eq:schmidt5}) and (\ref{eq:schmidt6}) separately, except for $|\tilde{p}_0\rangle\otimes|\tilde{n}_0\rangle$ and $|p_0\rangle\otimes|n_0\rangle$, as they both have the same parameters.
In this case $|\psi_c\rangle$ in Eq.~(\ref{eq:gradient}) represents $\lambda_0|\tilde{p}_{0}\rangle\otimes|\tilde{n}_{0}\rangle+\lambda_1|p_{0}\rangle\otimes|n_{0}\rangle$.
Once all the gradients are computed, the largest one determines which operator from the pool is chosen and in which circuit is implemented. 
To compute the energy, we extract the statevector from each circuit, make the corresponding tensor products between proton and neutron states, add them up as in Eq.~(\ref{eq:schmidt6}), and compute the expected value of the Hamiltonian with the obtained statevector.
The energy computation in an actual quantum computer could be implemented as proposed in~\cite{PerezMarquez2023}, with the exception of matrix elements of the Hamiltonian involving different product states.

We also notice that, considering the normalization of the truncated state, we only have one free parameter $\lambda$. We can include $\lambda$ as another parameter in the optimizer when classically finding the minimum of the energy surface.
In this work, we fix $\lambda$ to the value given by the Schmidt decomposition, to speed up testing of entanglement forging and focus on the VQE part.
Degenerate states are also related by symmetry. For the FH decomposition, Eq.~(\ref{eq:schmidt5}), the four degenerate states are related by spin and parity transformations. Therefore, only two independent product states and four circuits need to be optimized, for example $|l_{\uparrow\downarrow}\rangle\otimes|r_{\uparrow\downarrow}\rangle$ and $
|l_{\uparrow\uparrow\downarrow}\rangle\otimes|r_{\downarrow}\rangle$.
Similarly, the five degenerate product states in Eq.~(\ref{eq:schmidt6}) are related by parity. Slater determinants with  same angular momentum $j$ and opposite $m$ satisfy $|j,m\rangle=(-1)^{j-m}|j,-m\rangle$, with $|j,m\rangle\equiv a_{j,m}^\dagger|0\rangle$. We can then simulate only $|p_{-2}\rangle\otimes|n_{2}\rangle$, 
$|p_{-1}\rangle\otimes|n_{1}\rangle$, 
$|p_{0}\rangle\otimes|n_{0}\rangle$, and obtain $|p_{1}\rangle\otimes|n_{-1}\rangle$, $|p_{2}\rangle\otimes|n_{-2}\rangle$ through a parity transformation. Thus we need a total of four product states and eight circuits.

\subsection{Second layer of entanglement forging}
For the FH model, we do not consider a second layer of entanglement forging as there is no clear partition to take advantage of, and the effects of the barrier can be properly studied with just the first layer.

We can apply the same procedure to each of the circuits simulating states with only proton or neutron orbitals, and consequently use circuits with one fourth as many qubits as orbitals there are in the shell.
As a second decomposition we choose to split each proton and neutron partition into low and high energy subshells. In the case of the \emph{sd}-shell, the bottom half consists of the lowest subshell, $0d_{5/2}$, while the upper half includes subshells $1s_{1/2}$ and $0d_{3/2}$. For the \emph{pf}-shell, this second cut involves splitting the subshell $1p_{3/2}$ in half, as shown in Fig.~\ref{fig:shells}.

This division is useful to test our approach beyond one-cut, but it is not as good as the proton-neutron separation in terms of entropy. Ideally, physical systems that exhibit more sectors with low entanglement would be better suited for a larger amount of cuts, including divisions into $t$ circuits instead of successive cuts in $2$. Also, in contrast to the first decomposition, where we have degenerate Schmidt coefficients for product states with well defined spin $J_z$ and isospin $T_z$, in this second decomposition we are free to choose the particular distribution of product states. In this case we decompose each product state into states with a different particle number distribution between the bottom and top orbitals. For example, if we have two valence protons,
we can have both protons in the bottom subshells and none in the others, labeled as (2,0), one in each partition, (1,1), or both protons in the upper subshells, (0,2).
These are all the possibilities to distribute the protons in all the neon and titanium isotopes.
For ${}^{28}$Ne and ${}^{60}$Ti we have 10 and 18 valence neutrons respectively, implying two holes in both nuclei.
We can distribute these holes as (0,2), (1,1) and (2,0).

\begin{figure*}[!t]
     \centering
     \includegraphics[width=1\linewidth]{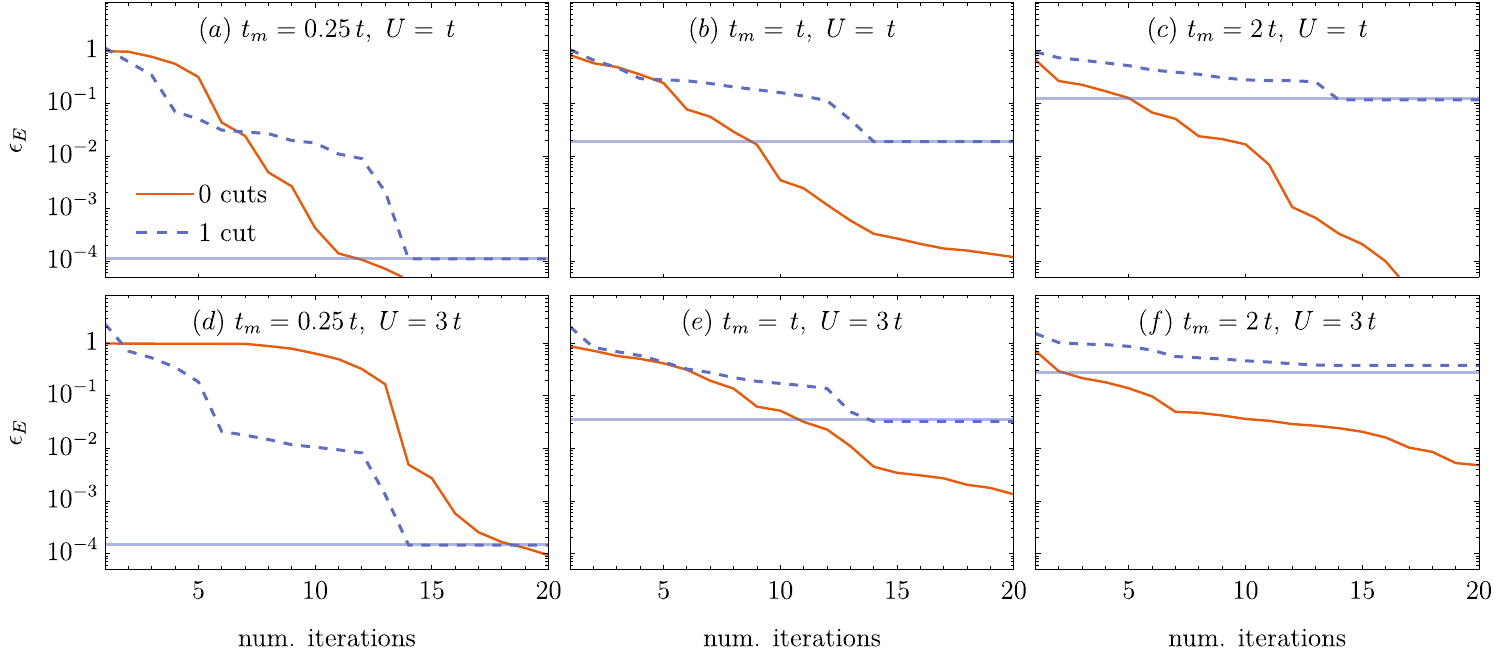}
     \caption{Relative errors in energy $\epsilon_E$ for the same FH simulations with interaction $U$, middle hopping $t_m$ pairs and "cuts" as shown in Fig.~\ref{fig:results_FH}. Horizontal lines mark the errors $\epsilon_E$ determined by the corresponding Schmidt decompositions of the given layer (lower blue line). These errors decrease substantially as $t_m$, and thus entanglement, diminish from right to left.
          } 
     \label{fig:FH_energy}     
\end{figure*}

To add only few more parameters to the ansatz we decompose each of the previous statevectors with well defined spin and isospin into only two terms with different particle distributions.
For the first  term in Eq.~(\ref{eq:schmidt6}), with $J_z^{(p)}=J_z^{(n)}=0$ and labeled $|\psi_{00}\rangle$, we consider two terms with a symmetric distribution of protons in the lower and higher energy orbitals, of $(2,0)$ and $(0,2)$. For the corresponding neutron states, the distribution of neutron holes is $(0,2)$ and $(2,0)$,
\begin{equation}
\begin{aligned}
|\psi_{00}\rangle\equiv&\,
\lambda_0|\tilde{p}_0\rangle\otimes|\tilde{n}_0\rangle
\\=&
    \,( b_1|p_{20}\rangle + b_2|p_{02}\rangle)\otimes
    ( b_1'|n_{20}\rangle + b_2'|n_{02}\rangle)
    \\=&\,
    \left(b_1|p_{2}^{(b)}\rangle\otimes |p_{0}^{(t)}\rangle + b_2|p_{0}^{(b)}\rangle\otimes|p_{2}^{(t)}\rangle\right)
    \\&
    \otimes
    \left(b_1'|n_{0}^{(b)}\rangle\otimes |n_{2}^{(t)}\rangle + b_2'|n_{2}^{(b)}\rangle\otimes|n_{0}^{(t)}\rangle\right).
\end{aligned}
\end{equation}
In the other five states, we assume a  distribution of $(2,0)$, $(1,1)$ for protons, and $(0,2)$, $(1,1)$ for neutron holes. That is, the most energetically favorable one.

Considering two terms also allows to not have to impose a normalization condition, since one of the coefficients is determined by the other. We only impose bounds on $b_1$, $b_2$ such that these coefficients can be normalized to $\lambda_0$ or $\lambda_1$, which are again fixed to the values given by the Schmidt decomposition. 
For example, $\sqrt{b_1^2+b_2^2}=\sqrt{\lambda_0}$,
and the same for the normalization of $|n_{0}\rangle$, $\sqrt{{b_1'}^2+{b_2'}^2}=\sqrt{\lambda_0}$
such that their product is normalized to $\lambda_0$. This also displays the training procedure for the coefficients that we propose for general use.

In Appendix~\ref{app:optimization} we show other more technical improvements in the optimization of ${}^{60}$Ti, which go beyond splitting the circuits into halves and fourths and which allow to speed up the simulation.

\section{Further details on the convergence of EDEF}
\label{app:cnots}

\begin{figure}[!t]
     \centering
     \includegraphics[width=1\linewidth]{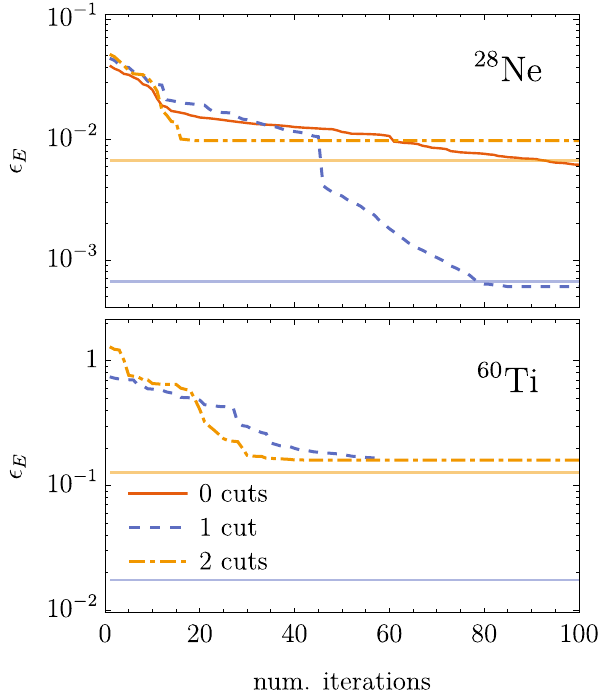}
     \caption{Relative errors in ground state energy for ${}^{28}$Ne (top) and ${}^{60}$Ti (bottom) as a function of the number of ADAPT-VQE iterations for 0-, 1-, and 2-cuts EDEF simulations.
     } 
     \label{fig:NeTi_energy}     
\end{figure}

In the main text, we show the convergence of the infidelity in two physical systems. This is an appropriate metric to test the method since the infidelity ties directly to the Schmidt decomposition. The training of the circuit, however, is based on an energy minimisation. For this reason, we also include and discuss here the relative errors in the energy, $\epsilon_E$,  as the iterations of the optimization progress. Namely, in Fig.~\ref{fig:FH_energy} we show the counterpart to Fig.~\ref{fig:results_FH} for the FH in terms of energy. We also show in Fig.~\ref{fig:NeTi_energy} the counterpart to Fig.~\ref{fig:netierrors} for the NSM. The notation of the figures follows that of the main text.

The two figures indicate that, in both systems, the behaviour of the relative energy error is qualitatively similar to that of the infidelity, already described in the main text. For the FH model, in particular, Fig.~\ref{fig:FH_energy} is also quantitatively similar to Fig.~\ref{fig:results_FH}. We find that EDEF outperforms standard ADAPT-VQE with the system settings of panel (d), with mild entanglement between the partitions. In all other cases, the EDEF minimisation saturates an energy bound calculated indirectly from the Schmidt decomposition within about $15$ iterations. This bound is not however a strict lower bound for the EDEF simulations. Indeed, panels (c) and (e) show examples where the bound is mildly improved by EDEF. 

For the NSM, the relative energy analysis in Fig.~\ref{fig:NeTi_energy} also shows  qualitative agreement to the infidelity results presented in Fig.~\ref{fig:netierrors} for $^{28}$Ne and $^{60}$Ti. Here, however, the quantitative discrepancies are more visible. First, the ADAPT-VQE energies for ${}^{28}$Ne decrease steadily with iterations, unlike the corresponding infidelities of  Fig.~\ref{fig:netierrors}, which increase at times. The rate of energy decrease for the full ADAPT-VQE simulation, however, is relatively slow: we require $100$ iterations to achieve a relative energy error of $\epsilon_E \approx 6 \times 10^{-3}$. In contrast, the 1-cut EDEF reaches a relative energy error which is about an order of magnitude smaller with only $80$ iterations (and far fewer resources). 
We stress again that relative energy errors given by the corresponding Schmidt decompositions, marked with horizontal lines, do not define a lower bound. In fact, for the ${}^{28}$Ne 1-cut EDEF simulation, the converged $\epsilon_E$ is slightly lower than its corresponding Schmidt decomposition $\epsilon_E$. In contrast, the 2-cut EDEF simulation reaches an error of only $\epsilon_E \approx 10^{-2}$, but does so within less than $20$ iterations. This error is somewhat larger than the bound suggested by the Schmidt decomposition, a result that is commensurate with the infidelity analysis of Fig.~\ref{fig:netierrors}. We take this as an indication of the limitation of the 2-cut approach, that cannot capture the residual entanglement of the second cut.

The results for $^{60}$Ti presented in the bottom panel of Fig.~\ref{fig:NeTi_energy} show a similar picture, although at a different level of quality (note the difference in $y-$axis between the two panels). As discussed in the main text, memory limitations preclude us from simulating the statevector for this isotope with full ADAPT-VQE. The 1- and 2-cut EDEF simulations start relatively far away from the final result (with $\epsilon_E \approx 1$). They subsequently decrease in energy steadily, at a rate that is comparable for the two simulations. This is at odds with the infidelity results of Fig.~\ref{fig:netierrors}, which indicates that the 2-cut infidelity decreases much faster than the 1-cut simulation. We note however that  the minimisation process is such that the infidelity and the energy lie relatively far from the Schmidt bounds. 

\begin{figure}[t!]
     \centering
     \includegraphics[width=1\linewidth]{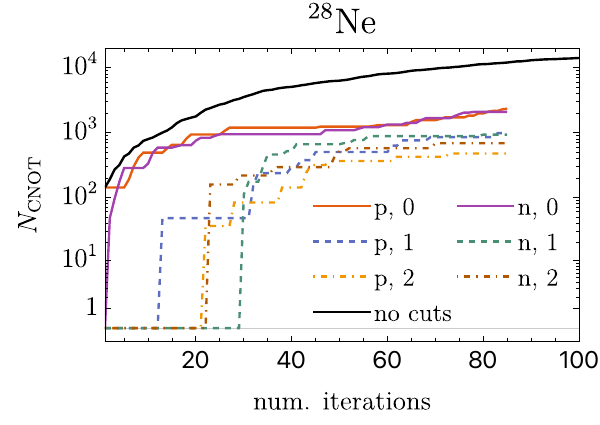}
     \caption{CNOT count  in each circuit for the simulation of ${}^{28}$Ne as function of iteration number. Full ADAPT-VQE results (black solid line) are compared to the $N_\text{CNOT}$ for each subcircuit of 1-cut EDEF simulation. We distinguish circuits with different third component $J_z$, see Eq.~(\ref{eq:schmidt6A}). 
     } 
     \label{fig:28necx01}     
\end{figure}

\begin{figure}[!t]
     \centering
     \includegraphics[width=1\linewidth]{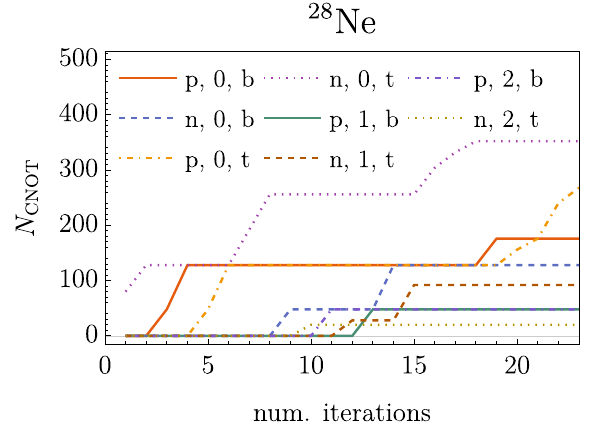}
     \caption{CNOT count  in each circuit for the simulation of ${}^{28}$Ne as function of iteration number. We show $N_\text{CNOT}$ for each subcircuit of a 2-cut EDEF simulation. 
     } 
     \label{fig:28necx02}     
\end{figure}

We also extend the amount of detail in the CNOT counts, $N_{\rm{CNOT}}$, by taking into account the effect of each subcircuit for 1-cut (Fig.~\ref{fig:28necx01}) and 2-cut (Fig.~\ref{fig:28necx02}) EDEF. We perform this analysis for the NSM only in the interest of brevity. Figure~\ref{fig:28necx01} shows the number of CNOTs for regular ADAPT-VQE (solid black line), together with $N_{\rm{CNOT}}$ for each of the circuits in the simulation of ${}^{28}$Ne with one layer of entanglement forging. First of all, we stress the fact that $N_{\rm{CNOT}}$ for all 1-cut EDEF circuits are notably lower throughout all the evolution. We find substantially different $N_{\rm{CNOT}}$ values depending on the state represented by the different circuits. The subcircuits associated to the first product states, $| p_0 \rangle$ and $| n_0 \rangle$, and the equivalent circuits for $| \tilde{p}_0 \rangle$ and $| \tilde{n}_0 \rangle$, that start from a different state but are otherwise identical, have a much larger $N_{\rm{CNOT}}$ count, as shown in the red and puple solid lines in Fig.~\ref{fig:28necx01}. In contrast, the states $| p_1 \rangle$, $| p_2 \rangle$, $| n_1 \rangle$ and $| n_2 \rangle$ are represented by much shallower circuits. In particular, these circuits are never selected by the VQA before the tenth iteration. The evolution of the CNOT count for these states is much more staggered. Each time an operator in the circuit is selected, the  $N_{\rm{CNOT}}$ count raises substantially with a single, relatively large step. The different staggering for each subcircuit suggests that each circuit is only picked up in the minimisation process at different stages. Clearly, in the initial stages of the minimisation the dominant product states with $\langle J_z \rangle=0$ are picked up, whereas states with lower Schmidt coefficients are only selected later in the minimisation process. In terms of total counts, within single-cut EDEF, we distinguish two scales: $N_{\rm{CNOT}}\simeq 2\times10^3$ for subcircuits simulating the first product state, with largest singular value, and $N_{\rm{CNOT}}\lesssim 10^3$ for the rest of subcircuit. 

We provide a similar analysis for the CNOT counts for all subcircuits with 2-cut EDEF for ${}^{28}$Ne in Fig.~\ref{fig:28necx02}. Here, we focus on the eight subcircuits which contain CNOT gates with different Schmidt coefficients,  belonging to proton and neutron partitions at the top or bottom of the single-particle spectrum. The number of CNOT gates in these 2-cut simulations is much smaller than the corresponding values of 1-cut simulations. The maximum number is well below $400$ across all the iterations, and corresponds to the dominant state in the neutron partition with single-particle energies at the top of the spectrum (eg $1s_{1/2}$ and $0d_{3/2}$). This leading circuit has $\langle J_z \rangle =0$ and is shown in a purple dotted line. The corresponding proton subcircuit (dash-dotted orange line) is the second most CNOT intensive circuit, reaching $N_{\rm CNOT} \approx 200$ at the end of the simulation. Neutron and proton bottom partitions (with $0d_{5/2}$ configurations)  end up the minimization with about $N_{\rm CNOT} \approx 100-150$. The remaining circuits correspond to smaller Schmidt coefficients and have less than $100$ CNOT gates by the time the simulation finishes. Overall, this bodes well with the idea that the subcircuits with largest Schmidt coefficients are picked up first, and more often, in the minimisation process.

\begin{figure}[!t]
     \centering
     \includegraphics[width=1\linewidth]{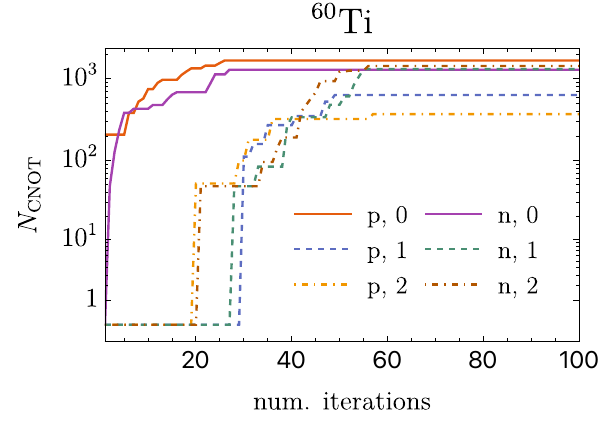}
     \caption{CNOT count in each circuit for the simulation of ${}^{60}$Ti as function of iteration number. We show $N_\text{CNOT}$ for each subcircuit of a 1-cut EDEF simulation.
     } 
     \label{fig:60tincx}     
\end{figure}

The simulations of ${}^{60}$Ti show similar results to ${}^{28}$Ne in terms of number of CNOTs for each subcircuit. We show in Fig.~\ref{fig:60tincx} the evolution of different subcircuit $N_{\rm CNOT}$ counts for the 1-cut simulation of ${}^{60}$Ti. ${}^{60}$Ti with one cut also shows a large difference between $N_{\rm{CNOT}}$ for $\langle J_z \rangle =0$ and $\langle J_z \rangle =1,2$ circuits. This is similar to the neon isotopes, where ADAPT-VQE chooses in most of the iterations to apply an operator to one of the circuits corresponding to the first product state. In fact, an operator is not applied to the second product states until the $20^\text{th}$ iteration. By the point that the simulation is converged, at iteration $50$, the proton and neutron circuits for the first product state have about $10^3$ CNOT gates each. For this neutron-rich isotope, however, the circuits associated to neutrons with $\langle J_z \rangle =1$ and $2$ have a similar number of gates, in spite of the relatively smaller Schmidt coefficient. The remaining proton subcircuits require about $500$ gates each. 

\begin{figure}[!t]
     \centering
     \includegraphics[width=1\linewidth]{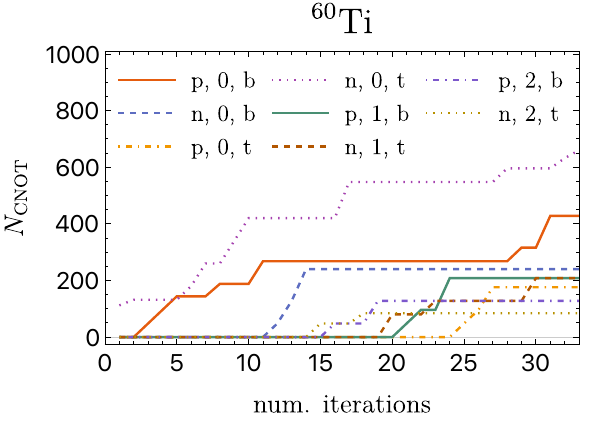}
     \caption{CNOT count in each circuit for the simulation of ${}^{60}$Ti as function of iteration number. We show $N_\text{CNOT}$ for each subcircuit of a 2-cut EDEF simulation. 
     } 
     \label{fig:60tincx2}     
\end{figure}

CNOT counts for ${}^{60}$Ti with two cuts are plotted in Fig.~\ref{fig:60tincx2}. The circuits with most CNOTs are much shallower than 1-cut ADAPT-VQE with the same number of iterations. In this case, the circuit with the largest $N_{CNOT}$ is the one for neutrons with $\langle J_z \rangle=0$ corresponding to the top subshells, which contains $660$ CNOTs. This is followed by the corresponding $\langle J_z \rangle=0$  proton circuit for the bottom partition, with over $400$ gates. Whereas for ${}^{28}$Ne, the $\langle J_z \rangle=0$ proton state for the top single-particle partition was the second subcircuit in terms of CNOTs, for ${}^{60}$Ti the same state ranks sixth in terms of CNOT counts. This is indicative of the individual complex pathways in the minimization process for different isotopes. 

\section{Results for a different isotope}
\label{app:26ne}

In the main text, we briefly mentioned how results of ${}^{28}$Ne compare to those of ${}^{26}$Ne. This isotope is an example with less particle/hole symmetry than the simulations in the main text, but for which the first layer of entanglement forging also performs well. We include here the details of such simulation for completeness. 

\begin{figure}[t]
     \centering
     \includegraphics[width=0.95\linewidth]{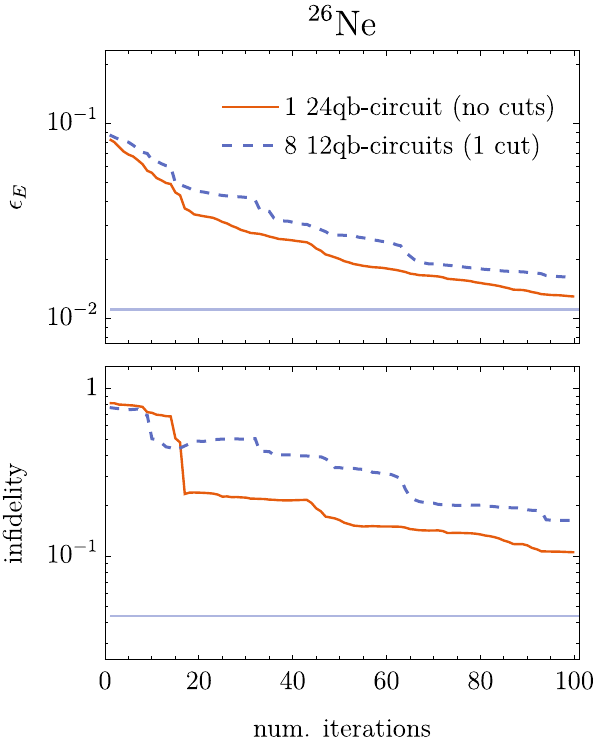}
     \caption{$\epsilon_E$ (top) and $I$ (bottom panel) for ${}^{26}$Ne as a function of the number iterations for ADAPT-VQE simulation (solid line) and 1-cut EDEF.
     } 
     \label{fig:ne26errors}     
\end{figure}

We summarize the EDEF results for ${}^{26}$Ne in Fig.~\ref{fig:ne26errors}. The top panel shows the relative energy energy as a function of the iteration number, in line with the results presented in Fig.~\ref{fig:NeTi_energy} for ${}^{28}$Ne. We plot results for 0- and 1-cut simulations with solid red and dashed blue lines, respectively. In this case, both lines have a very similar behaviour, decreasing gradually down to $\epsilon_E=0.013$ for full ADAPT-VQE and to $\epsilon_E=0.016$ for the 1-cut case. In terms of energy, ADAPT-VQE has not converged into a local minimum at the $100^\textrm{th}$ iteration. The 1-cut simulation follows a very similar trend and has not yet reached the bound associated to the Schmidt decomposition. 

The bottom panels focus on the evolution of the infidelity for the two methods. These should be compared to Fig.~\ref{fig:netierrors} in the main text. 
ADAPT-VQE here performs rather well, and the infidelity with respect to the ground state improves quickly between iterations $15$ and $20$, followed by a steady decrease afterwards. The infidelity reaches a value of $I=0.11$ at iteration $100$. In contrast, the evolution of the infidelity for the 1-cut simulation is somewhat more erratic. After an initial decrease, the infidelity increases between iterations $10$ and $30$, only to decrease later on at a rate that is more or less commensurate with the ADAPT-VQE results. The infidelity at the final step is $I=0.16$, a value that is still well above the corresponding Schmidt decomposition, $I^{(1)}=0.044$. Overall, the two pictures indicate that more iterations are required to reach a good representation of the ground state of ${}^{26}$Ne. The 1-cut EDEF approach may be able to reach this state with far fewer resources than the associated ADAPT-VQE simulation. 

\section{Optimization improvements}
\label{app:optimization}

The implementation of a variational algorithm such as ADAPT-VQE can fall short even when the ansatz and operator pool are adequate for the physical problem due to the necessity of training the parameters. Different implementations of the gradient descent work better for specific problems, but in general they may all run into resource problems as we try to deal with bigger systems. In this section, we present some optimization techniques that we implemented to increase the size of simulatable systems given a specific resource budget, which were specially useful to extend the simulation of the ${}^{60}$Ti nucleus into more layers.

First, we note that the more expensive step in our simulations is often the parameter tuning. However, we have heuristically found that the parameters  after an update $k$ are in general quite close to those of the previous step, $k-1$. Instead of optimizing over all parameters each time, one can find the optimal parameter only for the last operator that was added, $\theta_k$. We can then perform a complete update of all the parameters every $l$ rounds. Because now the operator with the biggest gradient is found with different parameters than the standard minimisation  (equivalent to $l = 1$), the list of chosen operators may change from the usual approach. This turns $l$ into a hyperparameter that may require tuning. In spite of these differences, we find that the precision of the training after the same number of steps is close with high probability, with very little dependence on the value of $l$. This is true even when the minimization employs a different set of chosen operators, consistent with the fact that the same ground state can be reached with different compositions of unitaries (due to the non-uniqueness of the unitary decomposition). While in general performance increases with $l$, one cannot make it too large because with fixed parameters the error can only be decreased by adding extra operators, and a deep circuit becomes exponentially expensive to simulate. Here, we choose $l=10$ for our ${}^{60}$Ti simulations.

In our entanglement forging approach to ADAPT-VQE, we find that the main circuit, the one with the leading coefficient on the Schmidt decomposition, is often prioritized by the operator choice. As we have mentioned, deep circuits are harder to simulate, so one can exhaust the resources available by always choosing the optimal operator even if some of the subleading circuits are almost empty. If we sometimes choose to place an operator in a circuit that is shallow (and therefore easy to compute), we can improve our error with a relatively inexpensive step (specifically, relative to the progress in the algorithm). In addition, since gradients are evaluated at a fixed point and the optimization happens over all parameter space, the error can decrease more on a single step by adding operators that have a non-optimal gradient. This makes this technique have an even bigger impact, as some of these steps will be more efficient and also more effective. We implemented this idea by excluding the ``full" circuit from the choice of operators. While this was done with an ad-hoc heuristic based on the optimisation duration in our specific setup, the approach could potentially be generalized by introducing a new hyperparameter $\varphi$.

\end{document}